\documentclass[aps, a4paper,12pt,reqno,superscriptaddress,nofootinbib]{revtex4}
\usepackage[centertags]{amsmath}
\usepackage{amsfonts}
\usepackage{amssymb}
\usepackage{amsthm}
\usepackage{newlfont}
\usepackage{stmaryrd}
\usepackage{mathrsfs}
\usepackage{euscript}
\usepackage{siunitx}
\usepackage{comment}
\usepackage{algorithm2e}
\usepackage{graphicx,subcaption}
\usepackage{enumitem}
\usepackage{natbib} 
\usepackage{braket}
\usepackage{bbm}
\usepackage{graphicx}
\usepackage{xcolor}
\usepackage{floatrow}
\usepackage{caption}
\usepackage{hyperref}
\usepackage{subcaption}
\usepackage{hhline}
\usepackage{hyperref}
\usepackage{cleveref}
\usepackage{amsmath}
\usepackage[bb=dsserif]{mathalpha}
\usepackage{bm}
\usepackage{tikz}
\usetikzlibrary{arrows.meta}
\usepackage{orcidlink}
\theoremstyle{plain}

\theoremstyle{definition}

\theoremstyle{remark}

\numberwithin{equation}{section}

\newcommand{\sech}{\mathrm{sech}}
 \let\be=\beta \let\de=\delta 
\let\ve=\varepsilon   
  \let\om=\omega

\let\bt=\beta

\newcommand{\opunit}{\text{1}\kern-0.22em\text{l}}

\DeclareMathAlphabet{\mathpzc}{OT1}{pzc}{m}{it}

\newcommand{\fig}{Fig.\;}

\newcommand{\id}{\textrm{d}}

\usepackage{floatrow}
\usepackage{caption}
\DeclareCaptionJustification{justified}{\leftskip=0pt \rightskip=0pt \parfillskip=0pt plus 1fil}

\numberwithin{equation}{section}

\usepackage{xcolor,soul}

\begin{document}

\title{Critical behavior of the driven Curie-Weiss model}
\author{Ruohan Xu \orcidlink{0009-0002-6505-9831}} \affiliation{School of Physics, Peking University, Beijing, 100871, China}
\author{ Faezeh Khodabandehlou \orcidlink{0000-0001-8114-6105}} \affiliation{Department of Physics and Astronomy, KU Leuven, Belgium}
\author{ Christian Maes\orcidlink{0000-0002-0188-697X}} 
\affiliation{Department of Physics and Astronomy, KU Leuven, Belgium}
\begin{abstract}
We complete the phase diagram of the macroscopic Curie-Weiss magnet in a time-periodic external field, as a function of temperature and driving parameters.  
%The nature of the phase transition changes from second order (close to equilibrium) to first order (for larger driving amplitude).
%There is a regime (only at large enough driving) where the paramagnetic and ferromagnetic phases coexist. 
There is a regime (large enough driving amplitude and frequency, at low temperatures) where stable paramagnetic and ferromagnetic phases coexist. In particular, we present a new detailed analysis of the (nonequilibrium) specific heat, diverging at the same critical inverse temperature $\beta_c$ as the magnetic susceptibility. 
%with critical exponents that are impossible in thermal equilibrium. For example, e
The new Curie temperature decreases with the driving, and we find critical exponent $\alpha=1$ for $\beta\downarrow \beta_c$, and $\alpha\simeq 0.86$ for $\beta\uparrow \beta_c$, even for small driving.  A Floquet analysis shows the nature of the criticality, which is dynamical, with implications that remain unseen and are mostly impossible when the system is in thermal equilibrium.
\end{abstract}

\maketitle
%\tableofcontents

\section{Introduction}
In sharp contrast with equilibrium statistical mechanics, the study of phase transitions and critical behavior in nonequilibrium steady states remains at a relatively early stage of development. Most available examples arise from exactly solvable models, such as the totally asymmetric exclusion process and related, often one-dimensional, systems \cite{krug,liggett,schutz}. These models, together with other stochastic interacting particle systems, including the Toom model, infection (contact) processes, and sand pile dynamics, have provided important insights, in particular by revealing novel types of phase transitions and by extending the relevant parameter space through the inclusion of a driving strength, \cite{zia}. At the same time, such models typically lack a clear thermodynamic interpretation and do not allow for systematic variation of standard physical parameters such as temperature or magnetic field. Even in cases where these parameters can be introduced via local detailed balance, as in driven lattice gases, analytical progress has proven extremely challenging, even at a perturbative level, and the characterization of critical behavior remains largely unresolved.\\

In the present work, we follow an approach analogous to the early study of critical phenomena in equilibrium systems, namely through mean-field theory. More specifically, we adopt a Landau-type description of Model-A relaxational dynamics for a single order parameter, considering the Curie–Weiss dynamics of the magnetization in a time-periodic magnetic field. Neither the use of a mean-field approximation nor the investigation of the driven Curie–Weiss model itself is new \cite{Chak1999, Berger2012, TomeOliveira1990, elena2, suzukikubo}. Numerical, analytical, and experimental studies have shown that this transition shares many features with equilibrium critical phenomena, such as scaling behavior, universality, and well-defined critical exponents \cite{TomeOliveira1990,Rikvold1994,Robb2007,Robb2008}. In particular, Gallardo \emph{et al.} derived a mean-field description of this transition and showed that the critical exponent for the divergent magnetic susceptibility are the same as in equilibrium mean-field theory, strengthening the link between equilibrium and nonequilibrium critical behavior in driven systems \cite{Berger2012}.\\
Our contribution is to introduce nonequilibrium calorimetry as an additional tool to elucidate the phase diagram and the associated critical behavior, following up on previous numerical work in \cite{elena2}. Remarkably, and supported by arguments based on Floquet theory, we find that the specific heat diverges at criticality. In contrast with the magnetic susceptibility, this behavior differs now fundamentally from the equilibrium case, where the specific heat exhibits a finite jump at the critical temperature, but does not diverge. We further show that the critical temperature decreases with increasing driving amplitude and can also be identified through the divergence of the magnetic susceptibility. Depending on the driving parameters, the transition may be of first or of second order, and we identify a regime in which the ferromagnetic and paramagnetic phases coexist.\\

In this way, we complete the phase diagram of the driven Curie–Weiss model, confirming and extending previous results reported in \cite{Berger2012}. Overall, the driven Curie–Weiss model displays a rich phase structure in which genuinely time-dependent features play a decisive role, particularly in determining the divergent behavior of the specific heat.\\
Other nonequilibrium versions of the Curie-Weiss model, taking multiple temperatures, have been studied in \cite{Beyen_2024, Ptaszy2025}.\\

\underline{Plan of the paper}:  We start in the next section with the definition of the driven Curie-Weiss dynamics.  Sections \ref{sphe} and \ref{ms} introduce the specific heat and the DC-magnetic susceptibility; we give a first exploration.  The  exploration of the phase diagram starts in Section \ref{Seccritical}.  Mostly using numerical work, we give evidence of a unique critical point in the sense that the divergences of specific heat and of magnetic susceptibility coincide with the boundary of uniqueness of phase.  We also give a good guess for the critical temperature.  By the end of Section \ref{phdi} we have completed the phase diagram where the main parameters are temperature, driving amplitude and driving frequency.  More analytic work starts with Section \ref{floan}.  We give boundaries to various phases and we establish the coexistence of ferromagentic and paramagnetic phases. That follows Floquet theory, which is continued in Sections \ref{divsp} and \ref{divmag} in the study of the divergences of specific heat and magnetic susceptibility. \\

\section{Driven Curie-Weiss dynamics}
\subsection{The model}
We consider the kinetic Ising model described within a mean-field approximation.  For a large number $N$ of spins, the Hamiltonian of the system is given by
\begin{align}\label{ha}
H = -\frac{J}{2N} \sum_{i,j=1}^N \sigma_i \sigma_j - h_0(t) \sum_{i=1}^N \sigma_i ,
\end{align}
where $\sigma_i = \pm 1$ are Ising spin variables and $J > 0$ is the coupling constant.  The time-dependent external field $h_0(t)$ is time-periodic, 
\begin{align}\label{delh}
h_0(t) =  \delta_0+A_0 \cos(\omega t),
\end{align}
where $\omega$ is the angular frequency and $A_0$ is the amplitude of the oscillating field, around a static field $\delta_0$.  Notice that $A_0$ and $\delta_0$ have units of energy; from now  we put $h(t)=h_0(t)/J$ and $A=A_0/J, \delta=\delta_0/J$.  The parameters $A,\omega$ determine the driving and take the system out of thermal equilibrium.\\
The order parameter is the magnetization $m^N =(\sigma_1+\ldots+\sigma_N)/N$ and we wish to understand its stationary behavior.  In the following we write $m(t)$ for the time-dependent magnetization (ignoring the capital $N$ in the subscript).\\

The time evolution of the system follows a relaxational stochastic dynamics \cite{Glauber1963,elena2,TomeOliveira1990}, with transition attempts occurring at rate $\nu$. In the macroscopic limit, $N\uparrow \infty$,  the time-evolution obtains an effective 
description in terms of  $m(t)$ alone: we get
a relaxation dynamics 
\begin{equation}\label{mag}
\frac{\id m(t)}{\id t} + \nu m(t) =\nu \tanh\,\big[\beta (m(t) + h(t))\big], 
\end{equation}
where $m = \langle \sigma_i \rangle$  (in the Curie-Weiss spin-flip dynamics for $N\uparrow \infty$) and   $\beta = J/(k_B\,T) $ is the dimensionless inverse temperature.  Note that we have chosen a specific choice of spin-flip rates for the dynamics; in general $\nu$ may still be a positive function of $m(t)$; see {\it e.g.} \cite{elena2} for examples.  For simplicity, here we set the time-scale by putting $\nu =1$ as reference frequency for $\omega$.\\
Evolution \eqref{mag} shows relaxation to a periodic solution, still denoted by $m(t)$, although at the critical point, the relaxation may be very slow --- see more below, in particular starting with Section \ref{floan}.\\

From equilibrium studies, we know that the specific heat and the magnetic susceptibility provide valuable information about the critical point. A natural question is therefore to study these quantities as well for nonequilibrium systems, and we need to define the specific heat and the magnetic susceptibility for the driven Curie-Weiss model corresponding to the steady solution of \eqref{mag}.

. \subsection{Specific heat}\label{sphe}
The specific heat of a driven Curie--Weiss model has been studied in \cite{elena2}, building on the nonequilibrium calorimetry presented in \cite{calo}. The central focus of these works is to characterize the thermal response of a driven nonequilibrium system under quasistatic perturbations. Here we introduce and study the specific heat for the driven Curie-Weiss dynamics \eqref{delh}--\eqref{mag} at $\delta=0$, extending \cite{Berger2012}.\\

In the macroscopic limit, the energy density of the system with Hamiltonian \eqref{ha} can be written as $E(t) = -J m^2(t)/2 - Jh(t)m(t)$. Differentiating with respect to time gives
\[
\dot{E}(t) = -J [m(t) + h(t)]\dot{m}(t) -J\dot{h}(t)m(t) = -J\,P(T) + J\,\dot{W}(t),
\]
where $J\,\dot{W} = -J\,\dot h\,m = m\,\id h_0/\id t$ is the external power exerted on the system, and $J \,P(t)$ is the dissipated power or heat flux to the  thermal bath at inverse temperature $\beta$. \\
In order to apply AC-calorimetry \cite{calo}, the thermal bath is subjected to a small, slow variation of the temperature. Specifically, we introduce a weak periodic modulation of the inverse temperature
\begin{equation}\label{tmod}
\beta_t = \beta \left(1 + \varepsilon \sin(\omega_B t)\right),
\end{equation}
where both \(\varepsilon\) and \(\omega_B \ll \omega\) are small parameters, ensuring a quasistatic regime.  The temperature modulation must be inserted in \eqref{mag}, giving
\begin{equation}\label{magt}
\frac{\id m(t)}{\id t} +  m(t) = \tanh\,\big(\beta_t (m(t) + h(t))\big). 
\end{equation}
Under the slow modulation \eqref{tmod} of the bath temperature, we expand the magnetization as
\begin{equation}\label{m1}
    m(t) = m_0(t) + \ve\, m_1(t) + \mathcal{O}(\ve^2),
    \end{equation}
where $m_1(t)=\frac{\partial m(t)}{\partial \ve}|_{\ve=0,\de=0}$ and $m_0(t)= m(t)|_{\de =0,\ve=0}$.  We use it for the dissipated power $J\,P(t)$ with
\begin{equation}\label{pm}
P(t) = \left(m(t) + h(t)\right)\, \frac{\id m(t)}{\id t}.
\end{equation}
Substituting \eqref{m1} into \eqref{pm}, we obtain
\begin{equation}
P(t) = P^0(t) + \varepsilon P^1(t) + \mathcal{O}(\varepsilon^2),
\end{equation}
where
\begin{align}
P^0(t) &= (m_0(t) + h(t))\,\frac{\id m_0(t)}{\id t},\label{p0} \\
P^1(t) &= m_1(t)\,\frac{\id m_0(t)}{\id t} + (m_0(t) + h(t))\,\frac{\id m_1(t)}{\id t} %Uing 
\label{p1}
\end{align}
At fixed $\beta$, {\it i.e.}, for $\varepsilon = 0$, the system dissipates heat into the bath with power $P^0(t)$, 
which, after transients, becomes periodic with frequency $\omega$. The correction term $P^1(t)$, 
depends on $\omega_B$ as well, so that the total power $P(t)$ is generally non-periodic (when $\omega$ and $\omega_B$ are non-commensurate).\\ 
Following the standard procedure of AC-calorimetry \cite{elena2, calo}, the specific heat $C = C(\beta, \omega,A)$ is extracted from the out-of-phase component of the excess $P^1(t)$.
\begin{equation}\label{hc}
    C=\lim_{\ve, \om_B \to 0, \de =0}\frac{\be}{\pi} \,\int_0^{2\pi/\omega_B} P^1(t)\,\cos(\omega_B t)\, \id t.
\end{equation}
and we wish to see its dependence on the driving parameters $A,\omega$ as well as to understand its divergence at the critical point.  A detailed analysis is deferred to the following sections; here, we first provide a glimpse of its behavior as a function of inverse temperature $\beta$ .\\

At low temperature,
\[
\tanh\,\bigl(\beta(m+h(t))\bigr) = \operatorname{sgn}\,\bigl(m+h(t)\bigr) + O\,\left(e^{-2\beta |m+h(t)|}\right).
\]
and hence, the evolution equation in \eqref{magt} becomes independent of $\beta$. Therefore, the limiting periodic solution $m(t)$ is itself independent of temperature. Equivalently, $\frac{\partial m(t)}{\partial \beta}\xrightarrow{\beta\uparrow\infty}0$.  Since $m_1(t)$ represents the linear response of the magnetization to a small modulation of the inverse temperature, it follows that $m_1(t)\xrightarrow{\beta\uparrow\infty}0$.
Recalling \eqref{p1}, every term in $P^1(t)$ is proportional either to $m_1(t)$ or to $\dot m_1(t)$. Hence, $P^1(t)\xrightarrow{\beta\uparrow\infty} 0$ and, in the limit of large $\beta$, its Fourier component at frequency $\omega_B$ vanishes as well. Therefore, for all driving parameters $\omega, A$,
\[
\lim_{\beta\to\infty} C(\beta,\omega,A)=0
\]
The vanishing of the heat capacity in the zero-temperature limit is consistent with a nonequilibrium extension of the Third Law of Thermodynamics, as discussed in \cite{mathnernst,jchemphys}.  \\

\begin{figure}[H]
 \centering
      \begin{subfigure}{0.49\textwidth}
         \centering
         \def\svgwidth{0.8\linewidth}        
        \includegraphics[scale = 0.37]{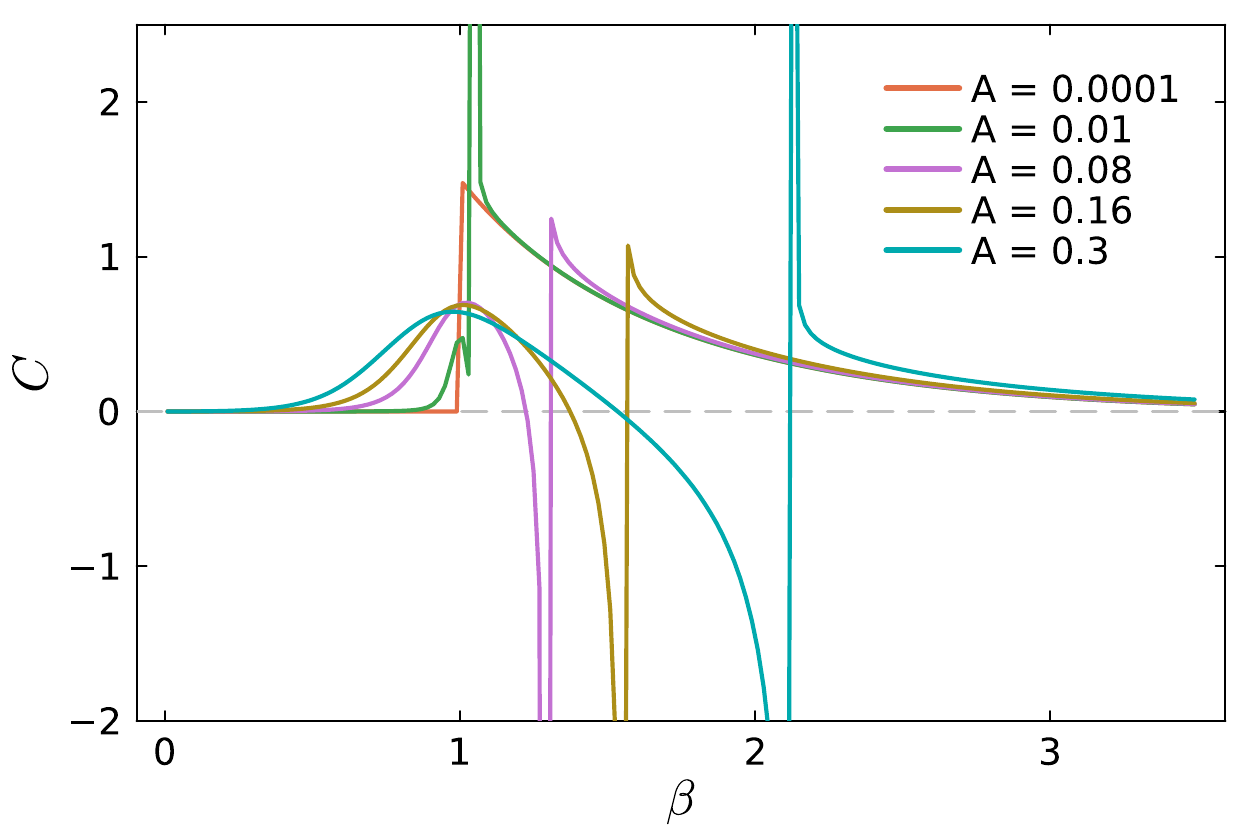}
        \caption{\small{$\om = 0.02$}}
     \end{subfigure}
     \hfill
     \centering
      \begin{subfigure}{0.49\textwidth}
         \centering
         \def\svgwidth{0.8\linewidth}        
        \includegraphics[scale = 0.37]{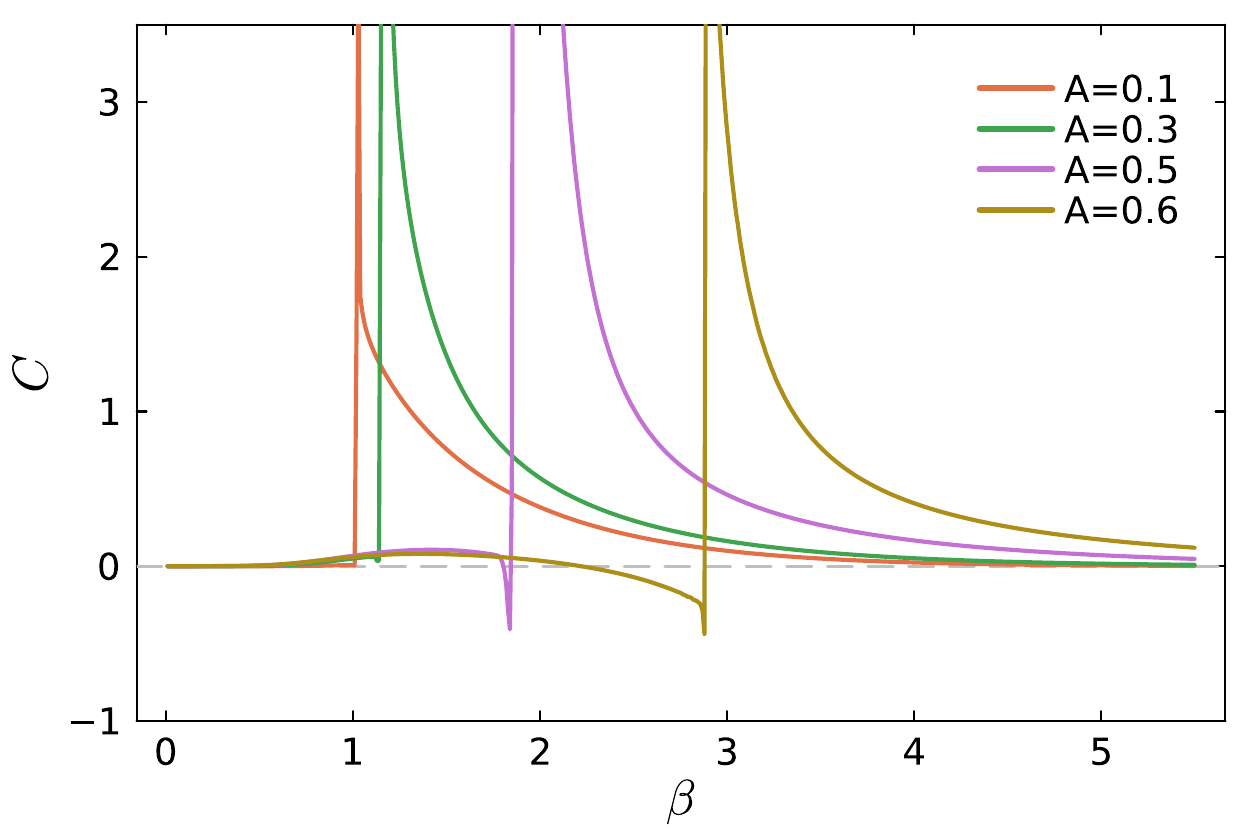}
        \caption{\small{$\om = 0.8$}}
     \end{subfigure}
\caption{\small{Specific heat as a function of inverse temperature $\beta$ for different values of $A$ and $\omega$. (a) reproduces Fig.~6 in \cite{elena2} for a small value of $\om$; (b) for larger $\om$ and $A$.} }  \label{CTgraph}
\end{figure}

 The specific heat \eqref{hc} is obtained by numerically solving \eqref{mag} for different values of $A$ and $\omega$, and is shown in Fig.~\ref{CTgraph}. 
\fig\ref{CTgraph} shows a divergence of the specific heat at a critical inverse temperature $\beta_c$. For $A=0$, the system is in equilibrium, with specific heat following the orange curve in \fig~\ref{CTgraph}(a) and (b), where the critical point is located at $\beta_c^\text{eq} = 1$ (Curie temperature). As the system is driven further out of equilibrium, the critical temperature (where the specific heat diverges) decreases. As we see in the next sections, that critical temperature is unique.  In Section~\ref{Seccritical}, we study in detail the behavior of the associated critical exponents.

\subsection{Magnetic susceptibility}\label{ms}
Besides the thermal response, an important probe of critical behavior is the magnetic susceptibility. Although the external magnetic field contains both static and oscillatory components, we focus on the DC-susceptibility and study its dependence on the driving amplitude $A$ and frequency $\omega$. The susceptibility measures the linear response of the magnetization to a small static perturbation $\delta$ of the external field, defined in \eqref{delh}, and is given by
\begin{equation}\label{xi}
    \chi    =    \left.    \frac{\partial \langle m\rangle}{\partial \delta}    \right|_{\delta=0},
    \qquad    \langle m\rangle    =    \lim_{\tau\to\infty}    \frac{\omega}{2\pi}    \int_{\tau}^{\tau+2\pi/\omega}
    m(t)\,\id t 
\end{equation}
where $\langle m\rangle$ denotes the asymptotic magnetization averaged over one driving period, and $m(t)$ is the solution of \eqref{mag} at fixed inverse temperature $\beta$. In general, the susceptibility depends on the temperature and on the driving parameters, $\chi=\chi(\beta,A,\omega),$ and we will characterize its critical behavior as well, similar to \cite{Berger2012}.\\

\begin{figure}[H]
 \centering
      \begin{subfigure}{0.49\textwidth}
         \centering
         \def\svgwidth{0.8\linewidth}        
        \includegraphics[scale = 0.37]{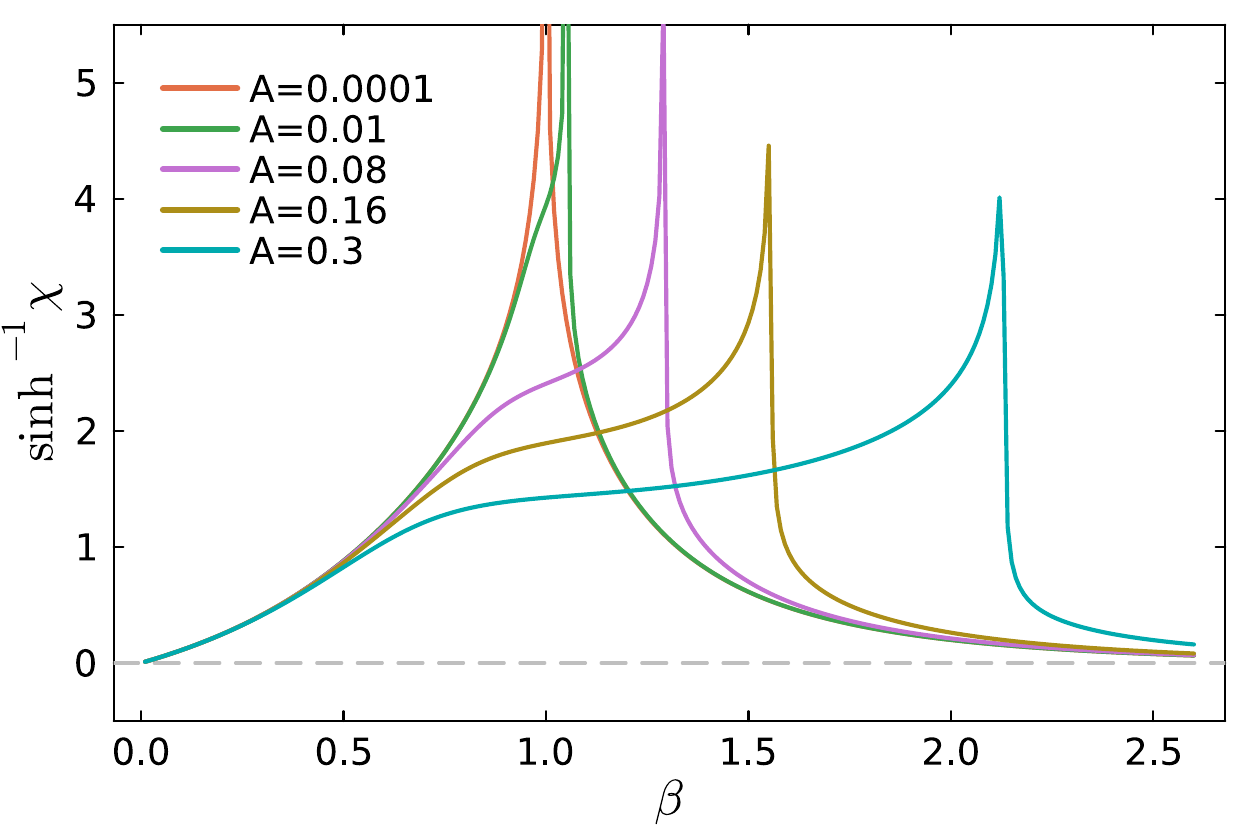}
        \caption{\small{$\om = 0.02$}}\label{chiTgrapha}
     \end{subfigure}
       \hfill
                 \centering
      \begin{subfigure}{0.49\textwidth}
         \centering
         \def\svgwidth{0.8\linewidth}        
        \includegraphics[scale = 0.37]{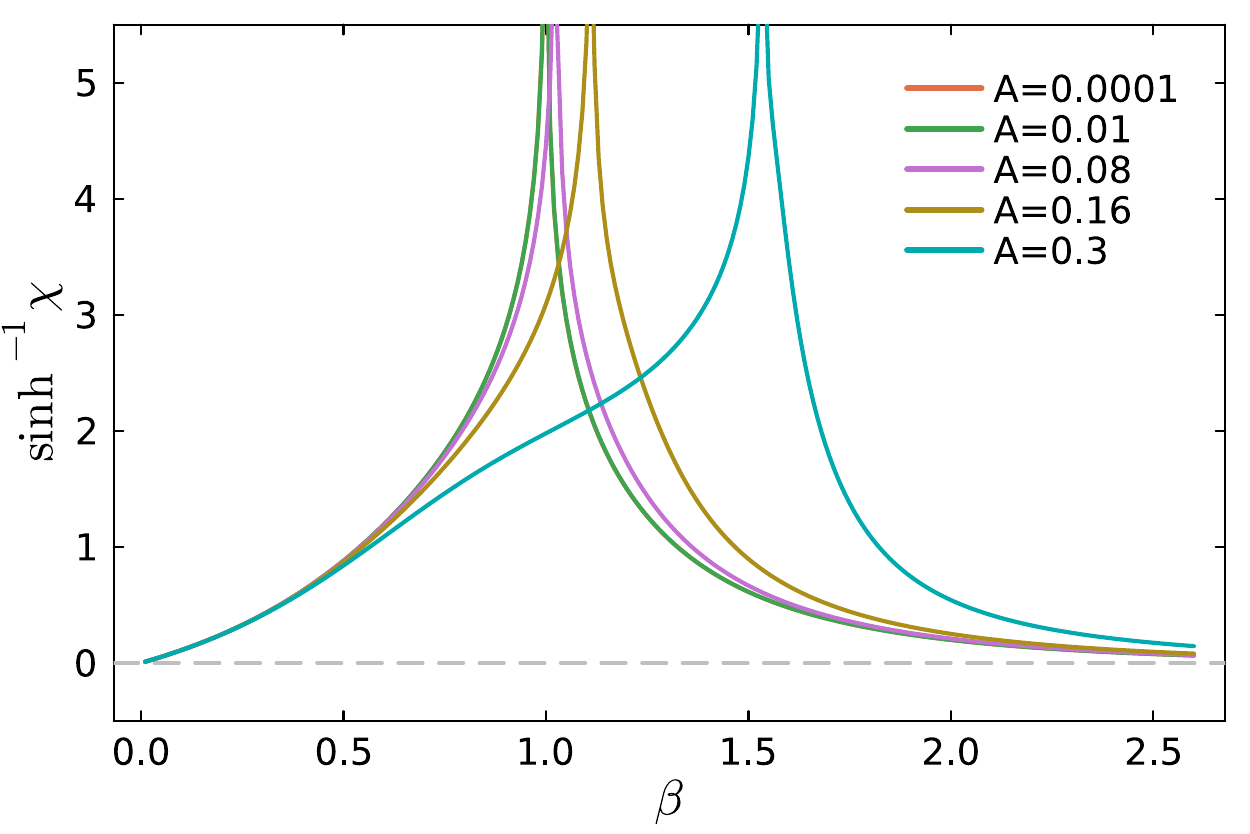}
        \caption{\small{$\om = 0.4$}}\label{chiTgraphb}
     \end{subfigure}
     \hfill
                 \centering
      \begin{subfigure}{0.49\textwidth}
         \centering
         \def\svgwidth{0.8\linewidth}        
        \includegraphics[scale = 0.37]{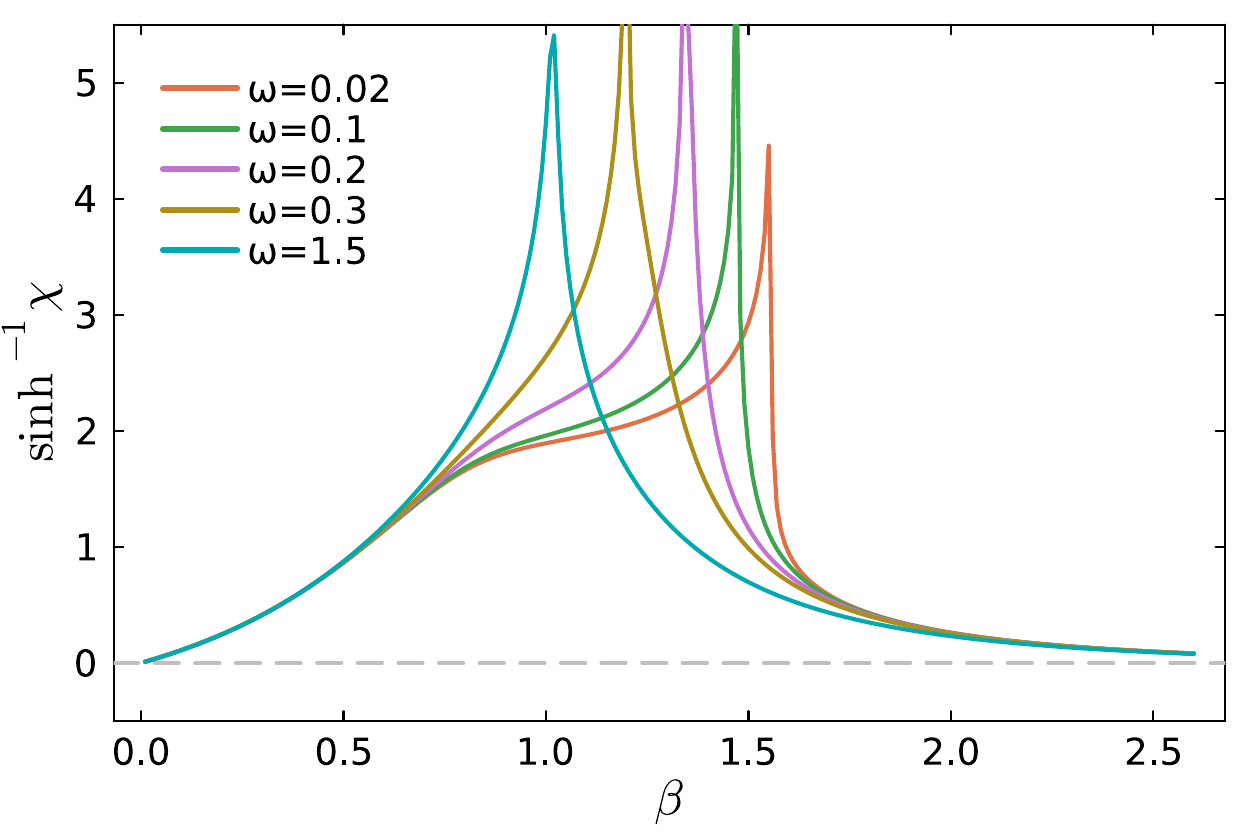}
        \caption{\small{$A=0.16$}}\label{chiTgraphc}
     \end{subfigure}
     \hfill
\caption{\small{$\sinh^{-1} \chi$ as function of $\beta$, for different values of $A$ and $\om$.} }  \label{chiTgraph}
\end{figure}

As a first illustration, Fig.~\ref{chiTgraph} shows the susceptibility as a function of $\beta$ for several values of $A$ and $\omega$. Since $\chi$ grows rapidly near the critical point, we plot $\operatorname{sinh}^{-1}(\chi)$ to improve visibility.
For small $A$ in \fig\ref{chiTgrapha}  and \fig\ref{chiTgraphb}, the magnetic susceptibility remains close to the equilibrium case. Still, for large values of $\omega$, the spins no longer have sufficient time to align, and as shown in \fig\ref{chiTgraphc}, the susceptibility approaches the equilibrium curve even for larger $A$.  Roughly speaking then, larger $A$ with small $\omega$ appears to bring the system truly away from equilibrium, while small $A$ stands for the close-to-equilibrium regime.\\

To compute $\chi$, we expand the magnetization for small $\delta$,
\begin{equation}
    m(t)=m_0(t)+\delta\,m_1(t)+O(\delta^2)
\end{equation}
Substituting this expansion into \eqref{mag} and using
\begin{align}
\tanh\,\big[\beta(m_0(t)+\delta (m_1(t)+1)&+A\cos (\omega t))\big]=\tanh\,\big[\beta(m_0(t)+A\cos (\omega t))\big]\\
&\quad +\delta\,\beta (m_1(t)+1)\,\sech^2\,\big[\beta(m_0(t)+A\cos (\omega t))\big]+O(\delta^2)\nonumber
\end{align}
we obtain the equations 
\begin{align}\label{m1m0dot}
\frac{\id m_0(t)}{\id t}&=\tanh\,\big[\beta(m_0(t)+A \cos (\omega t))\big]-m_0(t),\\[1ex]
\frac{\id m_1(t)}{\id t}
&=(m_1(t)+1)\Big[\beta\,\sech^2\,\big(\beta(m_0(t)+A \cos (\omega t))\big)\Big]-m_1(t) \label{m1m0dot1}
\end{align}
Since $m_1(t)=\left.\partial m(t)/\partial\delta\right|_{\delta=0}$, the DC-susceptibility is obtained by averaging the linear response over one period,
\begin{equation}
\chi=\lim_{\tau\to\infty}\frac{\omega}{2\pi}\int_{\tau}^{\tau+2\pi/\omega}m_1(t)\,\id t
\end{equation}

\section{Criticality at small driving amplitude}\label{Seccritical}
There are three ways to identify the criticality for the driven Curie-Weiss model: in the previous sections, we have introduced the thermal response (specific heat) and the magnetic response (DC-susceptibility); the third is via the analysis of the large-time behavior of the magnetization $m(t)$ and its dependence on $(\beta, A,\omega)$. In the present section, we limit ourselves to numerical and analytic explorations for small enough driving amplitude $A$. In Section\ref{phdi}, we show that the three approaches yield fully consistent results. \\

\begin{figure}[H]
   \centering
      \begin{subfigure}{0.49\textwidth}
         \centering
         \def\svgwidth{0.8\linewidth}        
    \includegraphics[scale=0.38]{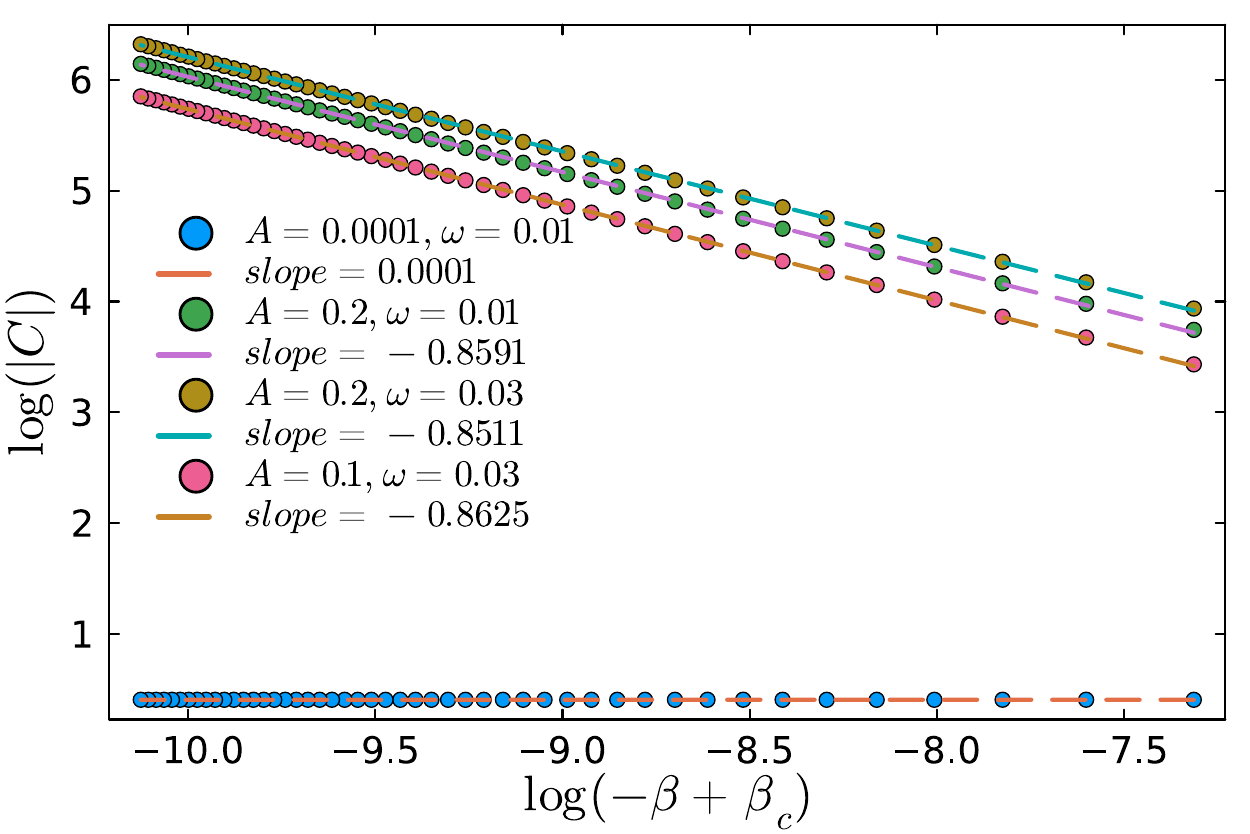} 
    \caption{$\beta< \beta_c$}\label{Ccriticalexfr}
     \end{subfigure}
     \hfill
     \centering
      \begin{subfigure}{0.49\textwidth}
         \centering
         \def\svgwidth{0.8\linewidth}        
        \includegraphics[scale = 0.38]{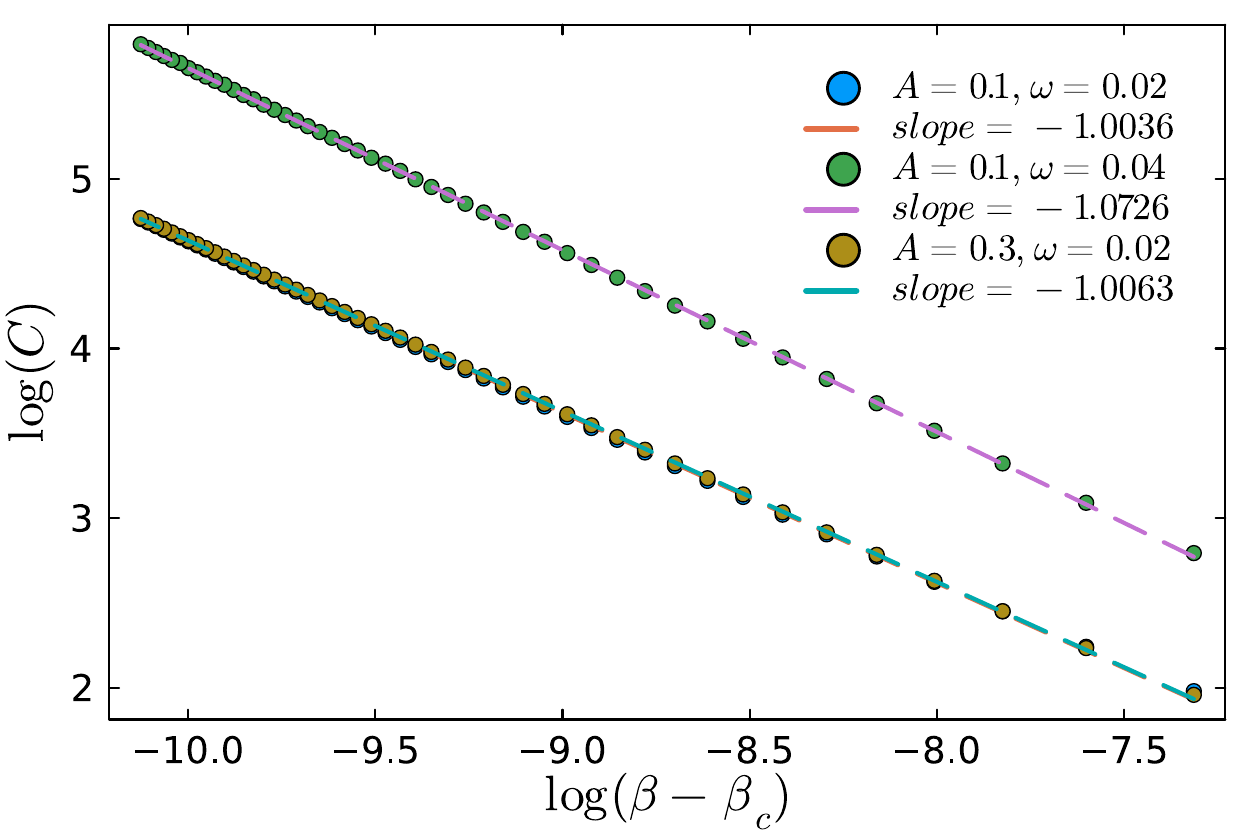}
        \caption{$\beta>\beta_c$}\label{Ccriticalexpar}
     \end{subfigure}
    \caption{\small{Critical behavior of the specific heat for different values of $A$ and $\omega$ in the vicinity of $\beta_c$. Scatter points represent the values of $\log |C|$ in (a) and $\log C$ in (b),  the dashed lines denote linear fits used to determine the critical exponent. (a) The fitted lines have slopes close to $-0.86$ for $\beta<\beta_c$, the blue scattered data are for equilibrium $A=0.0001$. (b)  The fitted lines have slopes close to $-1$ while $\beta>\beta_c$. }}
    \label{Ccriticalex}
\end{figure}

We begin by examining the divergence of the specific heat $C$ for different values of $A$ and $\omega$ in order to determine the critical inverse temperature $\beta_c$. As already illustrated in Fig.~\ref{CTgraph}, the equilibrium specific heat (in orange) exhibits a different behavior on the two sides of $\beta_c=1$.  Here also,  the behavior of the heat capacity is investigated in the vicinity of $\beta_c$, approaching the critical point from both the ferromagnetic, Fig.~\ref{Ccriticalexfr}, and paramagnetic phases, Fig.~\ref{Ccriticalexpar}.\\
In Fig.~\ref{Ccriticalexfr}, $\log |C|$ is plotted as a function of $\log(-\beta+\beta_c)$ for $\beta \uparrow \beta_c$ with $\beta < \beta_c$. The fitted slope on the side $\beta< \beta_c$ where $C$ is negative, yields slightly smaller values, around $0.86$.  While the critical exponent for $\beta\downarrow \beta_c$ differs from the value for $\beta \uparrow \beta_c$.   As shown in Fig.~\ref{Ccriticalexpar}, the fitted line has a slope close to $-1$. Thus, the specific heat exhibits a power-law divergence characterized by the critical exponent $\alpha \approx 1$ when the critical point is approached from the ferromagnetic side. We note that critical exponents $\alpha>1/2 $ are empirically rare in equilibrium.  \\

As a summary, 
\[
C(A,\omega;\beta) = K(A)\,(\beta-\beta_c)^{-1}, \beta >\beta_c,\quad
C(A,\omega;\beta) = K(A,\omega)\,(\beta_c-\beta)^{-0.86}, \beta < \beta_c
\]
However, this behavior is for the second-order transition where $A$ is sufficiently small; we see in Section \ref{phdi} how that changes for increasing $A$  where a first-order transition occurs.

\begin{figure}[H]
    \centering 
    \includegraphics[width=0.55\textwidth]{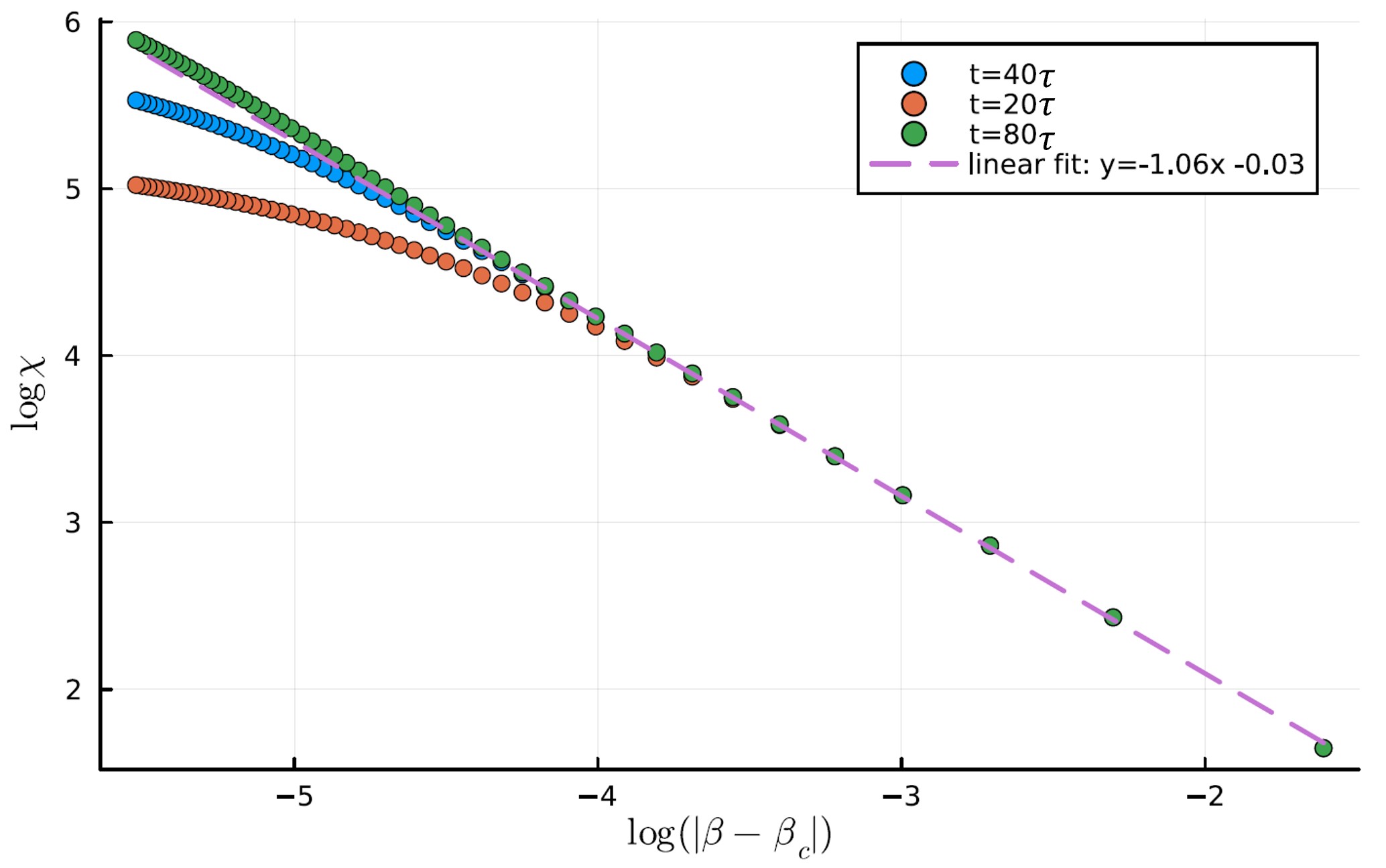} 
    \caption{\small{Critical behavior of $\chi$ for $A=0.1$ and $\omega=0.3$. The time $t$ is that of averaging as multiple of the period $\tau=2 \pi /\omega$.}}
    \label{int2}
\end{figure}

The same inverse critical temperature $\beta_c$  is found from the divergence of $\chi$.  In \fig \ref{int2}, we plot $\log \chi$ as a function of $\log|\beta - \beta_c|$ for different averaging times.   The fitted slope indicates that the critical exponent of the susceptibility is close to one, consistent with the value obtained for the specific heat.   Moreover, the plot shows that near the critical point, the relaxation becomes slower, requiring longer averaging times to obtain reliable results.

\begin{figure}[H]
   \centering
      \begin{subfigure}{0.49\textwidth}
         \centering
         \def\svgwidth{0.8\linewidth}        
    \includegraphics[scale=0.37]{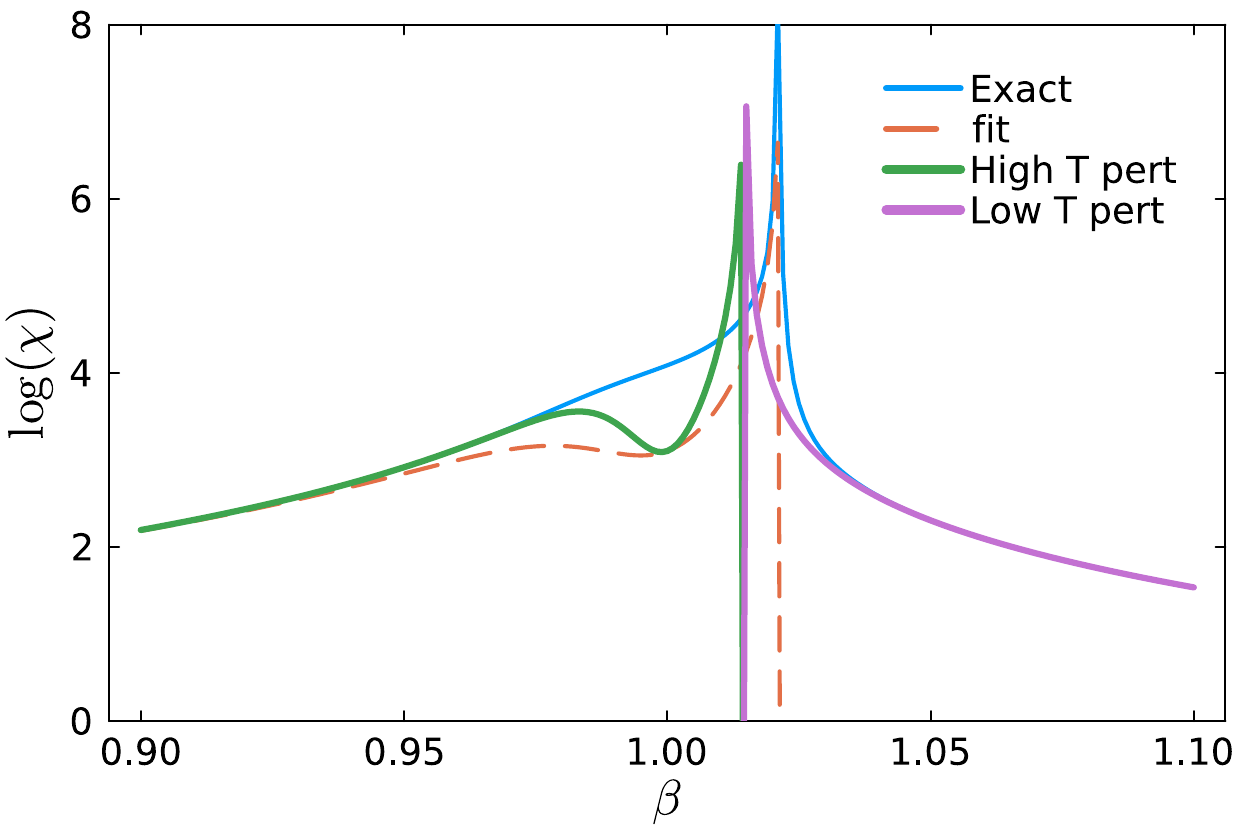} 
    \caption{\small{ $A=0.003 $ and $\om=0.01$. }}\label{fitta}
     \end{subfigure}
     \hfill
     \centering
      \begin{subfigure}{0.49\textwidth}
         \centering
         \def\svgwidth{0.8\linewidth}        
        \includegraphics[scale = 0.37]{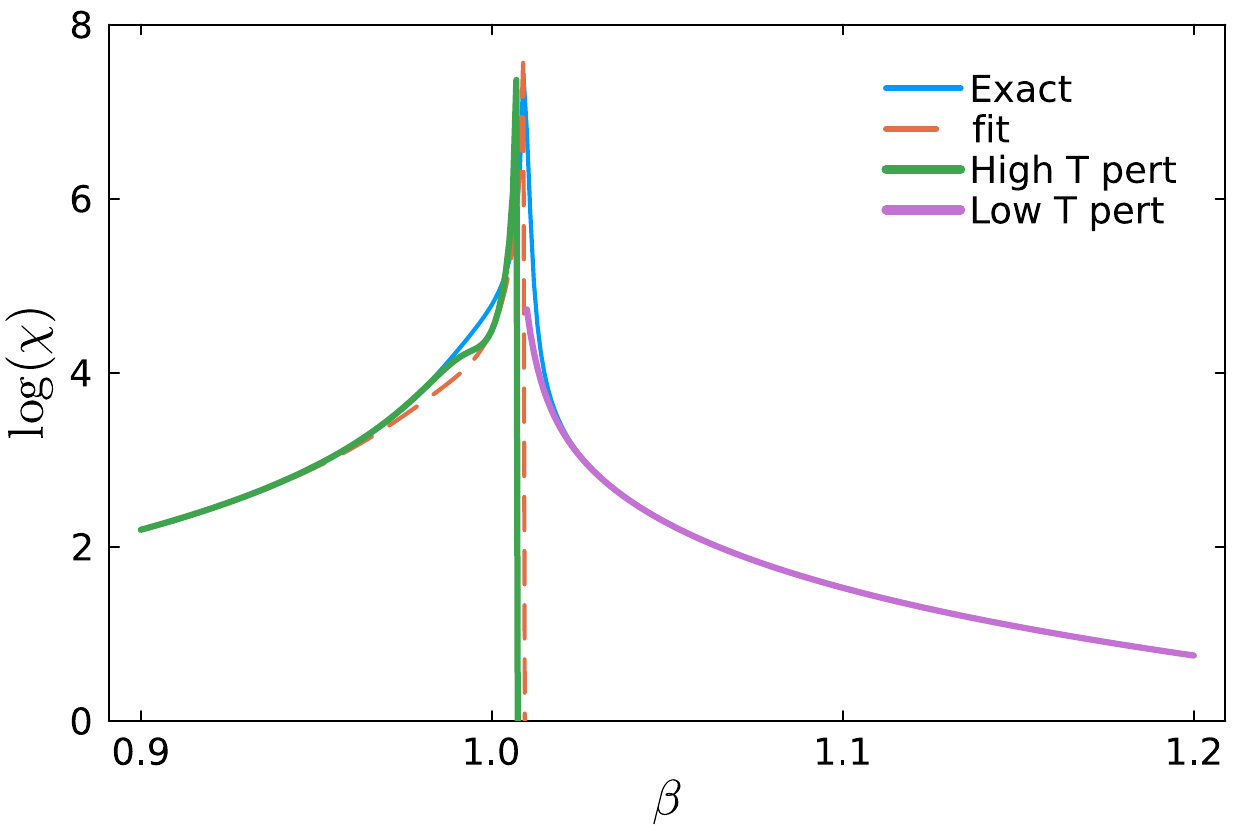}
        \caption{\small{$A=0.0015$ and $\om=0.01$}}\label{fittb}
     \end{subfigure}
    \caption{\small{Critical behavior of $\chi$, predicted by formula \eqref{strange},  high-temperature approximation \eqref{highfit} and low-temperature approximation \eqref{lowfit}, all for  small driving $A$.}}\label{fitt}
\end{figure}

\begin{figure}[H]
   \centering
      \begin{subfigure}{0.49\textwidth}
         \centering
         \def\svgwidth{0.8\linewidth}        
    \includegraphics[scale=0.37]{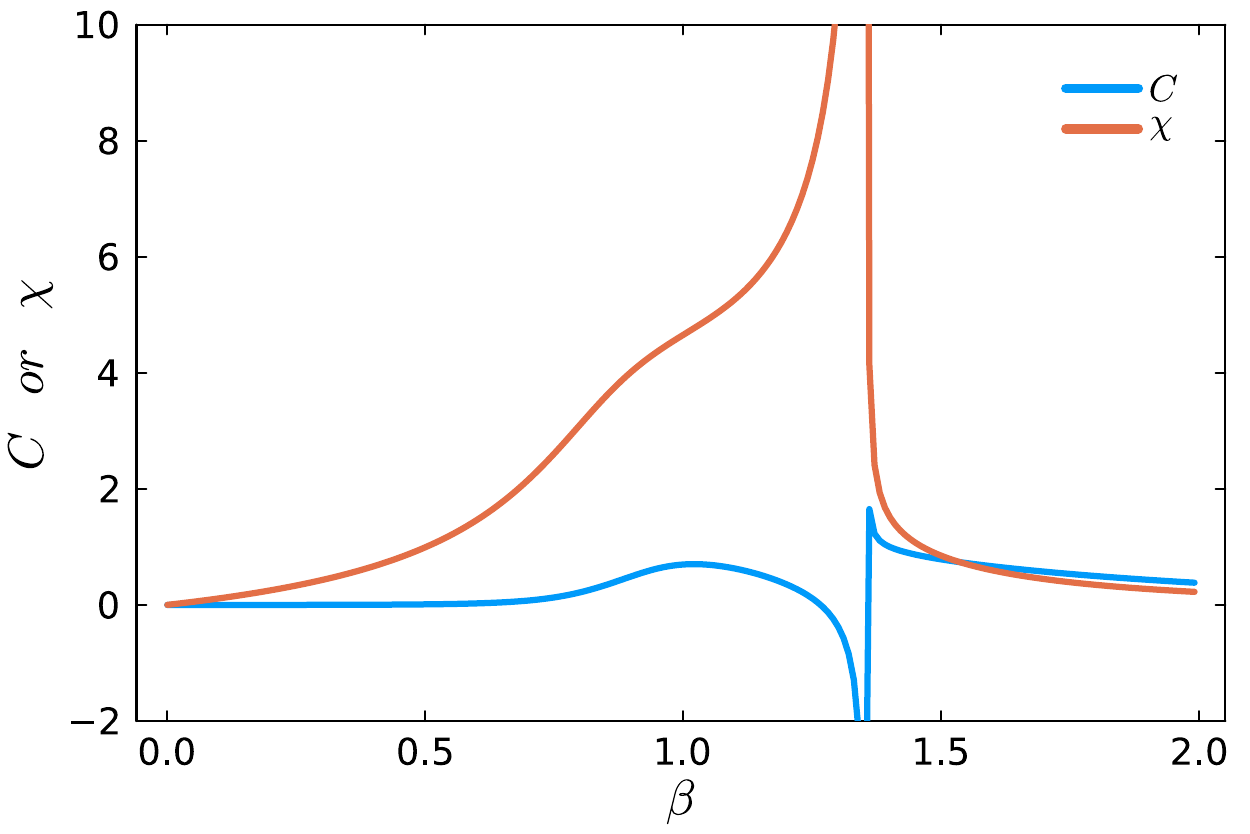} 
    \caption{\small{ $A=0.1, \,  \om=0.05$. }}
     \end{subfigure}
     \hfill
     \centering
      \begin{subfigure}{0.49\textwidth}
         \centering
         \def\svgwidth{0.8\linewidth}        
        \includegraphics[scale = 0.6]{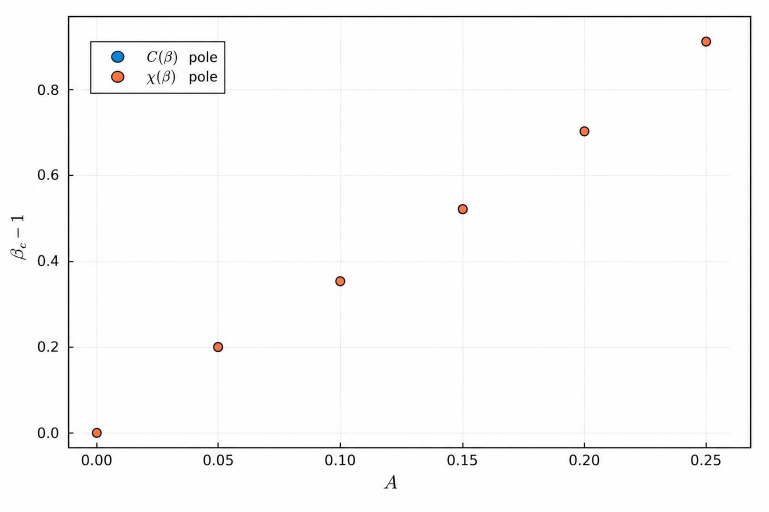}
        \caption{\small{$\om=0.02$}}
     \end{subfigure}
    \caption{\small{(a) Specific heat $C$ and susceptibility $\chi$ as  function of temperature. (b) Shift of the critical inverse temperature from its equilibrium value, $ 1$, as a function of the amplitude $A$. The specific heat $C$ and susceptibility $\chi$ yield overlapping estimates of the critical point for each value of $A$. }}
    \label{CCcriticalex1}
\end{figure}

\begin{figure}[H]
   \centering
      \begin{subfigure}{0.49\textwidth}
         \centering
         \def\svgwidth{0.6\linewidth}        
    \includegraphics[scale=0.28]{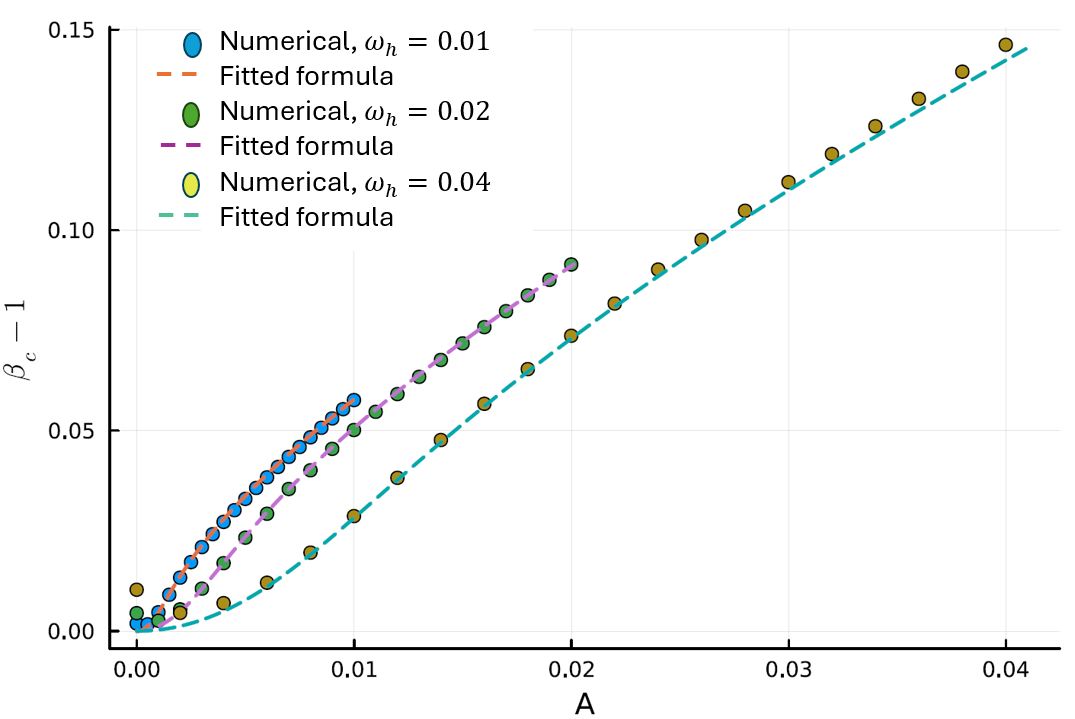} 
    \caption{\small{}}    \label{fig:ct_graph2a}
     \end{subfigure}
     \hfill
     \centering
      \begin{subfigure}{0.49\textwidth}
         \centering
         \def\svgwidth{0.6\linewidth}        
        \includegraphics[scale = 0.13]{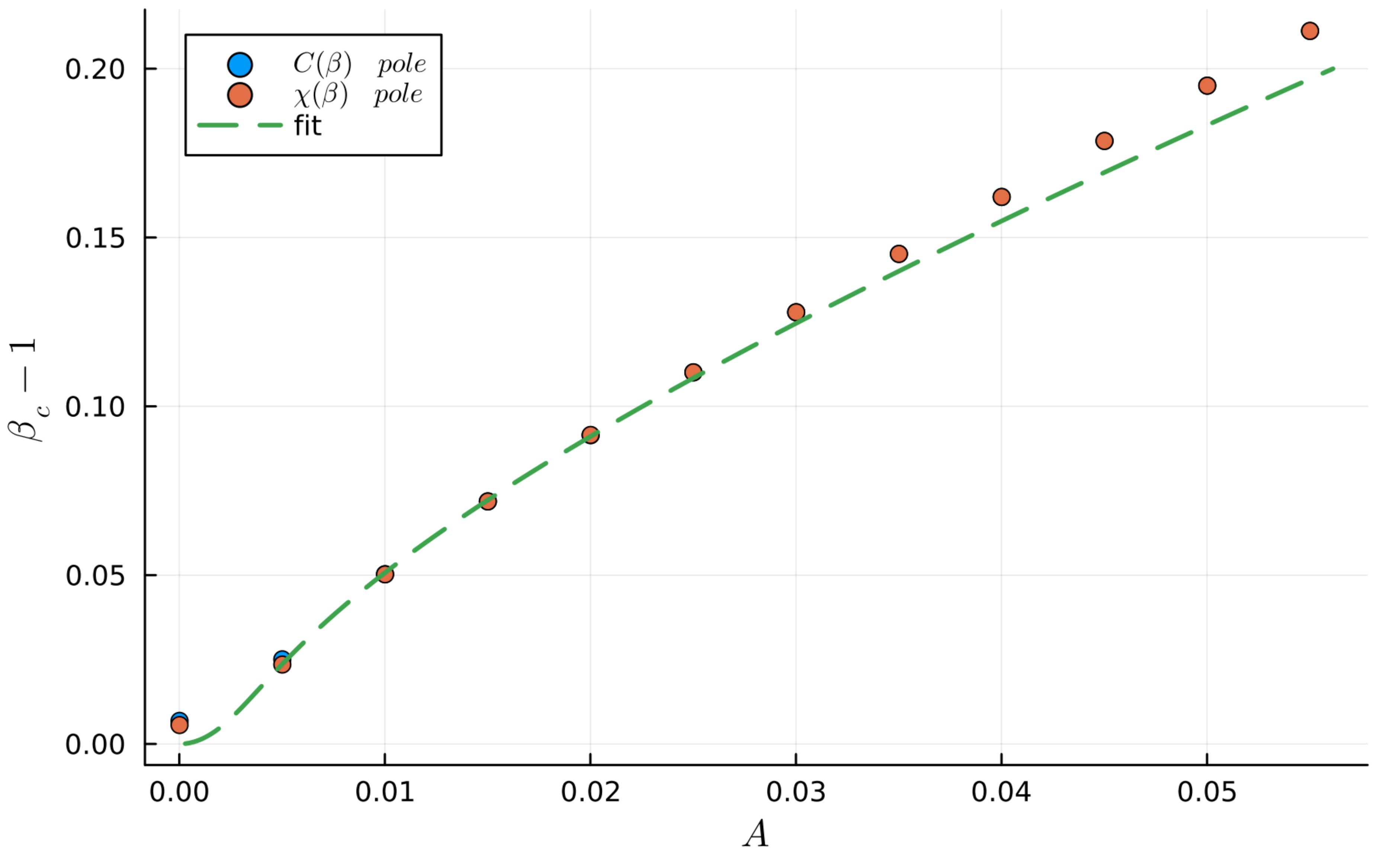}
        \caption{\small{$\omega_h=0.02$}} \label{fig:ct_graph2b}
     \end{subfigure}
\caption{\small{Behavior of the critical temperature as a function of the amplitude $A$. The data points are fitted using Eq.~\eqref{strange}. Figure (b) shows that the peak points of heat capacity and sustainability are mostly overlapped.}}
\label{fig:ct_graph2}
    \label{CCcriticalex}
\end{figure}

The magnetic susceptibility  for high temperature ($\beta \to 0$) is calculated in Appendix \ref{suscrit}, giving
\begin{align}\label{highfit}
    \chi \approx \frac{1}{b + \frac{A^2/2}{\om^2+b^2}}
\end{align}
where $b= \beta^{-1}-1 $ . By scaling that high-temperature formula to fit the critical point, we suggest that near-criticality
\begin{equation}\label{strange}
\chi^{-1} \approx \frac{b}{2} + \frac{\frac{A^2}{4}}{\om^2+\left(\frac{b}{2}\right)^2}
\end{equation}
We do not have a precise derivation of \eqref{strange}, but our guess is unexpectedly accurate; see Figs.\ref{fitt}--\ref{fig:ct_graph2}.\\
In the same Appendix \ref{suscrit}, we prove that  for the low-temperature regime,
\begin{align}\label{lowfit}
\chi = -g_1 + \frac{g_1(1+d)}{d-\Sigma}
\end{align}
where $g_1 = \bt \tanh'(\bt m_s)$, $d=1-g_1$, and $m_s$ is the solution of 
$m_s = \tanh(\bt m_s)$,
\begin{equation}
\Sigma = \frac{A^2 d}{g_1(1+d)}\left[\frac{\alpha^2+\gamma^2}{4}\left(\frac{g_3}{d}+\frac{g_2^2}{d^2}+\frac{2g_2^2}{d^2+\omega^2}\right)+\frac{g_2^2\,\omega\,\alpha\gamma}{d(d^2+\omega^2)}\right]
\end{equation}
for $g_2 = \bt^2 \tanh''(\bt m_s)$, $g_3=\bt^3 \tanh'''(\bt m_s)$.\\
In Fig.~\ref{fitt}, the above formulas are tested to demonstrate their accuracy in predicting the critical temperature. The fitted lines closely follow the main curve in the vicinity of the critical temperature, indicating excellent agreement.

As explored in Fig.~\ref{CCcriticalex1}, the susceptibility exhibits a divergence at the same critical point identified from the specific heat,  providing complementary evidence of the system’s critical behavior. We observe that  the approximate relations are working better for smaller $A$.

The fitting formula \eqref{strange} is also very accurate for predicting the value of the critical temperature in the driving region $A,\omega \ll 1$; see Fig.~\ref{fig:ct_graph2}. Near the critical point, the susceptibility diverges, and therefore the denominator in \eqref{strange} approaches zero. On the other hand, $b$ is very small, so that $b^2 \approx 0$. Consequently, we obtain $b=-\frac{A^2}{2 \omega^2}$, which yields  
\begin{equation}\label{effb}
\beta_c = 1 + \frac{A^2}{2\om^2}
\end{equation}

\section{Increasing the driving and full phase diagram}\label{phdi}

\subsection{Solving the magnetization dynamics for larger driving}
\begin{figure}[H]
 \centering
      \begin{subfigure}{0.49\textwidth}
         \centering
         \def\svgwidth{0.8\linewidth}        
        \includegraphics[scale = 0.37]{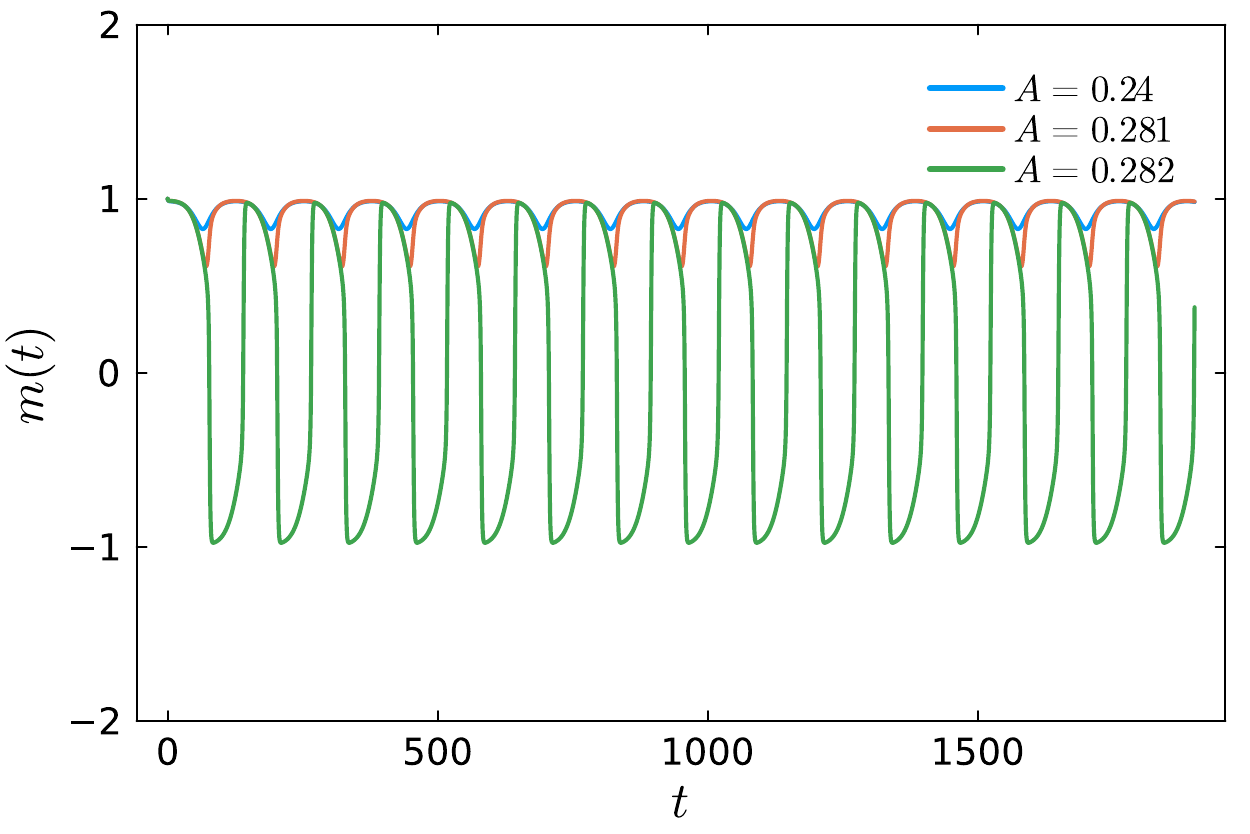}
        \caption{} \label{fig:ct_graph3}
     \end{subfigure}
     \hfill
     \centering
      \begin{subfigure}{0.49\textwidth}
         \centering
         \def\svgwidth{0.8\linewidth}        
        \includegraphics[scale = 0.37]{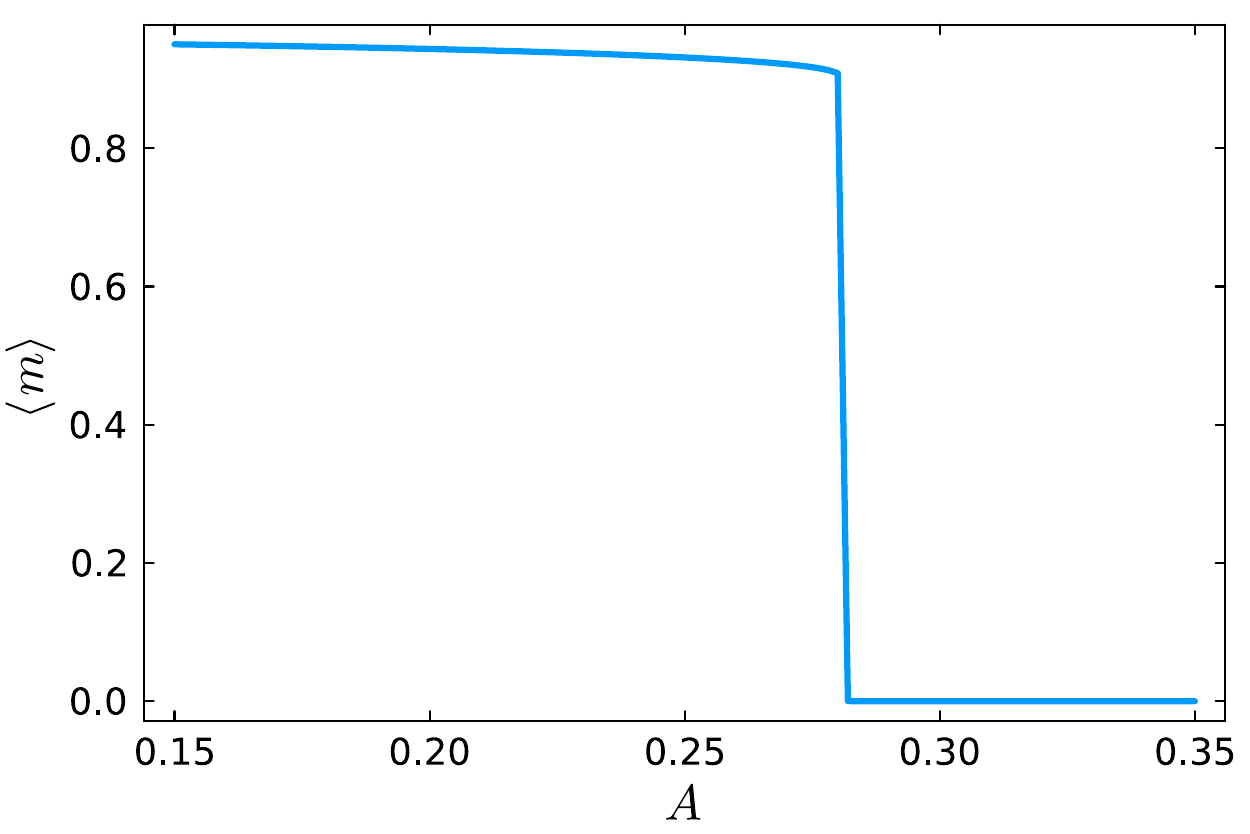}    \label{disc2}
        \caption{}
     \end{subfigure}
\caption{\small{First-order transition near $\beta=2$ and $\omega=0.05$ at $A\approx 0.281$. (a) $m(t)$ changes from a positive periodic solution to oscillating around zero. (b) Mean magnetization drops to zero.}  }  \label{phaseorder}
\end{figure}

We also look directly at the solution of \eqref{mag}. If the temperature is low and $\omega \ll 1$, we notice a sharp transition from a periodic positive solution to a periodic solution around zero, as we increase $A$ over the (same) critical point (as found above); see Fig.~\ref{phaseorder}.
Looking at the mean (time-averaged) magnetization $m = \langle m \rangle$, we observe a first-order phase transition.
 
\begin{figure}[H]
 \centering
      \begin{subfigure}{0.49\textwidth}
         \centering
         \def\svgwidth{0.8\linewidth}        
        \includegraphics[scale = 0.37]{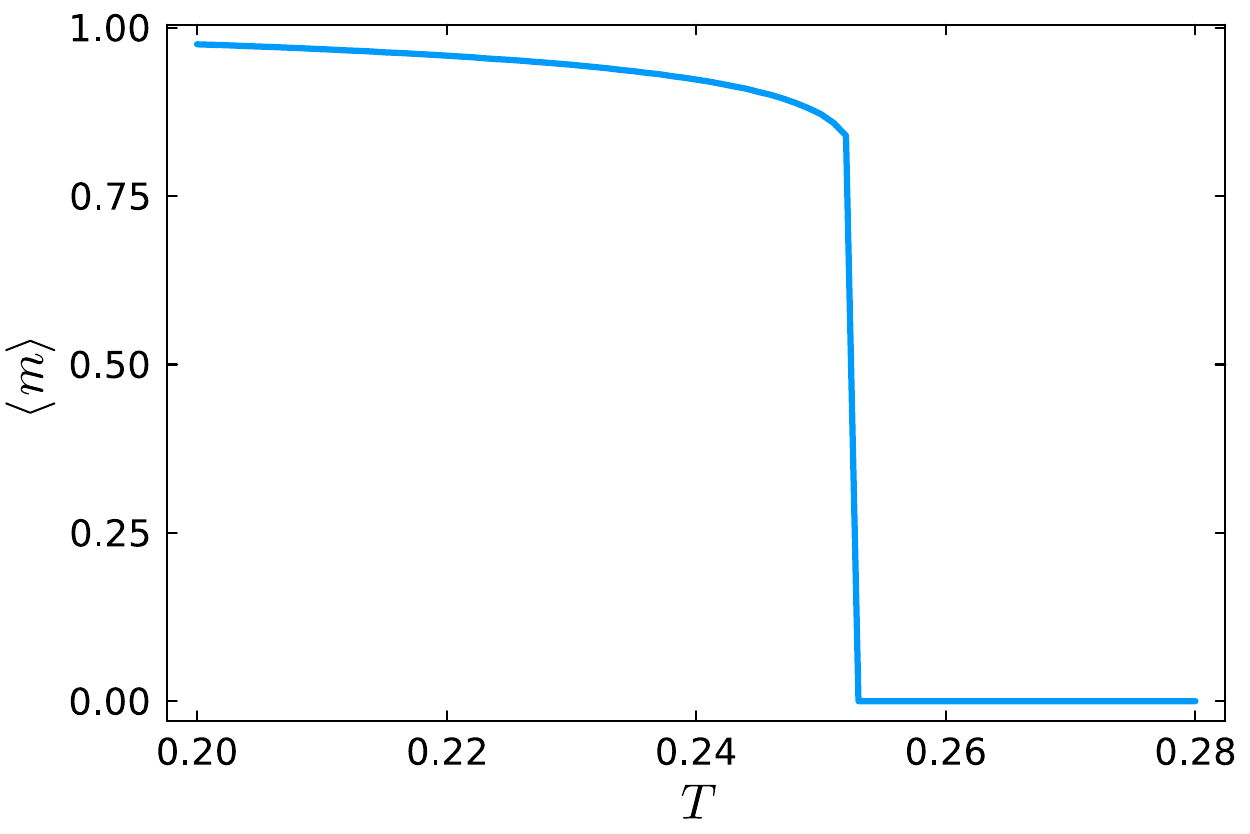}
        \caption{\small{Large $A=0.7$.}}
     \end{subfigure}
     \hfill
     \centering
      \begin{subfigure}{0.49\textwidth}
         \centering
         \def\svgwidth{0.8\linewidth}        
        \includegraphics[scale = 0.37]{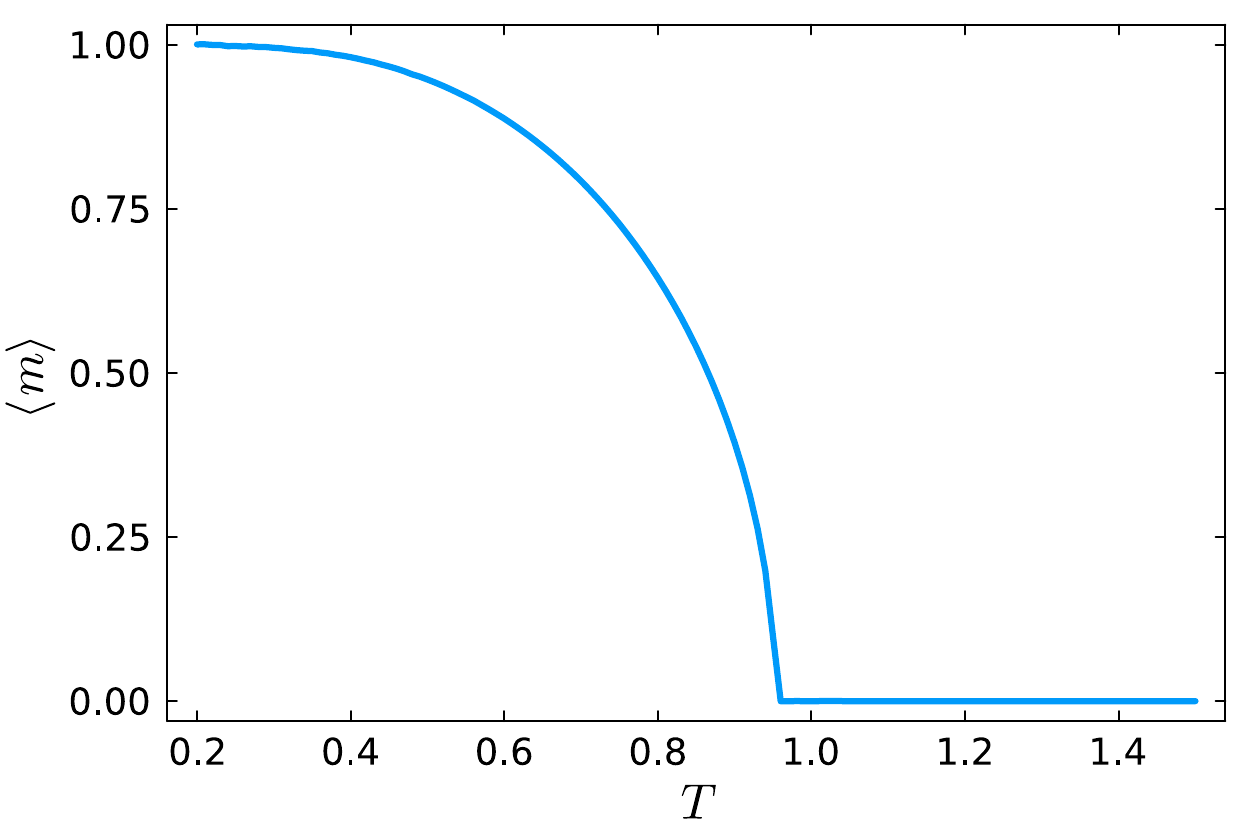}
        \caption{\small{Small $A=0.2$.}}
     \end{subfigure}
\caption{\small{Phase transitions at $\omega=0.9$. (a) First-order phase transition for large $A$ at low temperatures starting from $m(0)=0$.(b) Second-order phase transition for smaller $A$ at higher temperatures. } }  \label{phaseorder2}
\end{figure}

Similarly, for $\omega=0.9$ we see in \fig \ref{phaseorder2} a first-order transition at low temperatures at large amplitude ($A=0.7$), whereas for smaller amplitude ($A=0.2$) the transition becomes second-order and takes place at higher temperatures. 

\subsection{Phase diagram}
\begin{figure}[H]
 \centering
      \begin{subfigure}{0.49\textwidth}
         \centering
         \def\svgwidth{0.8\linewidth}        
        \includegraphics[scale = 0.38]{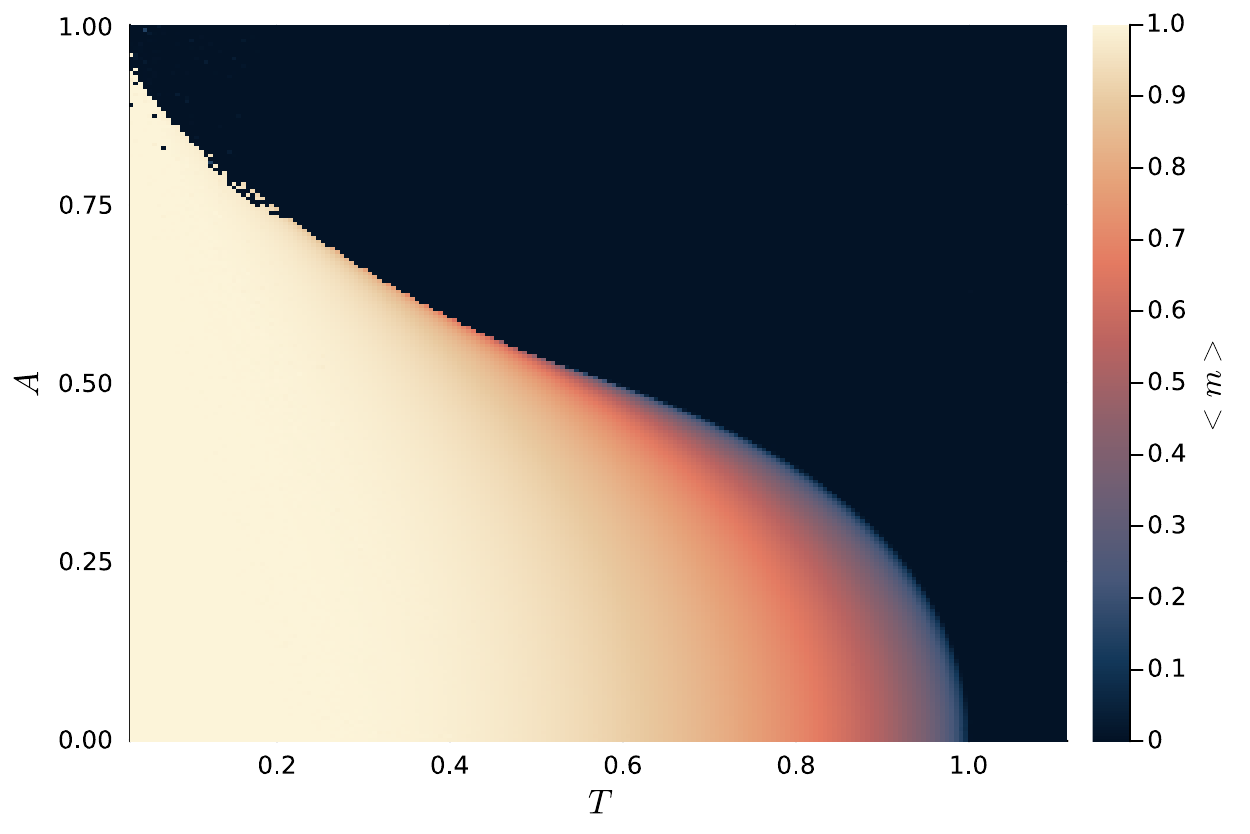}
        \caption{\small{}}\label{phaseAMa}
     \end{subfigure}
     \hfill
     \centering
      \begin{subfigure}{0.49\textwidth}
         \centering
         \def\svgwidth{0.8\linewidth}        
        \includegraphics[scale = 0.38]{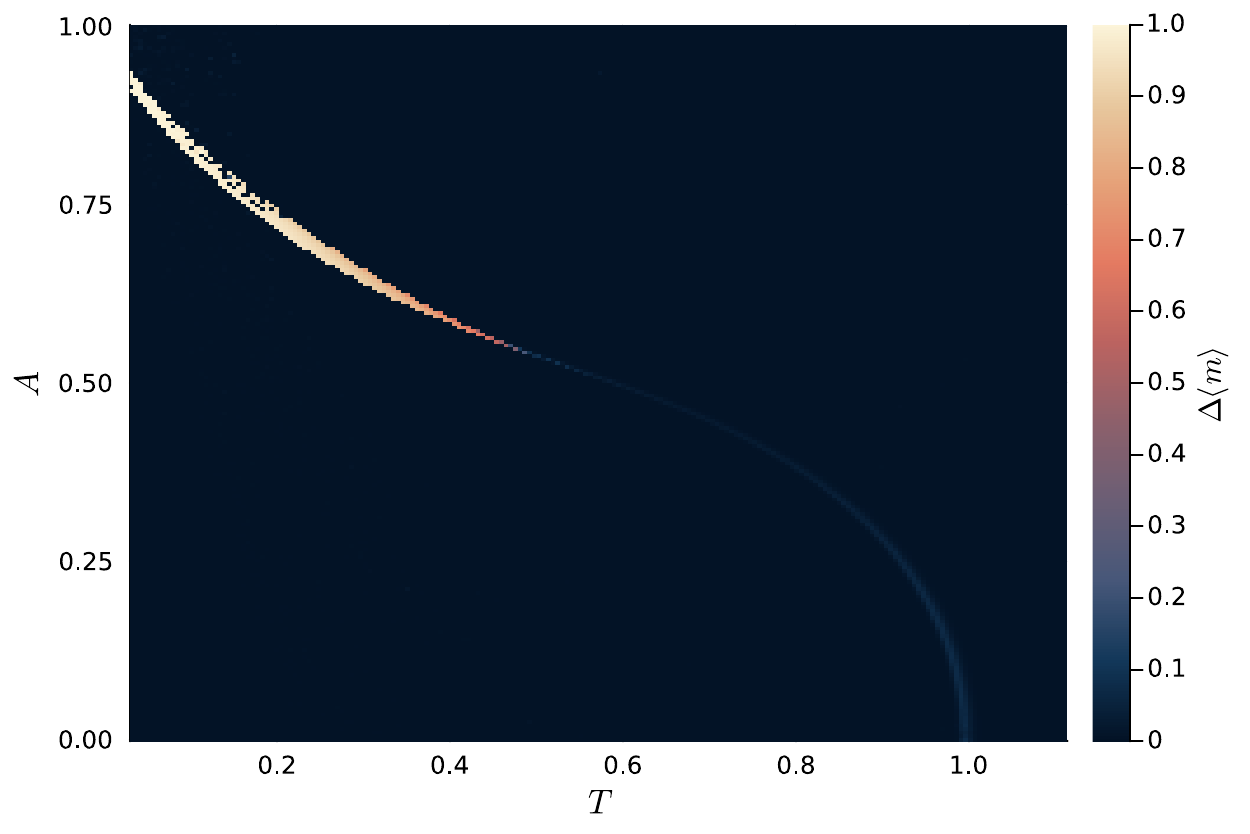}
        \caption{\small{}}\label{phaseAMb}
     \end{subfigure}
\caption{\small{  Phase diagram of the driven Curie--Weiss model for $\omega=0.9$.  (a) The diagram is obtained using ferromagnetic initial conditions 
$m_0=1$.  For small $A$ the transition appears continuous (second order), while for large $A$ it is discontinuous (first order).  (b) The coexistence region is revealed by comparing the long-time average magnetization $\Delta\langle m\rangle=\langle m\rangle_{m_0=0.99}-\langle m\rangle_{m_0=0.1}$.    } }  \label{phaseAM}
\end{figure}

By examining the magnetization for different values of driving amplitude and temperature, we obtain the phase diagram shown in Fig.~\ref{phaseAM}; it is fully comparable with the plot in \cite{Berger2012} (although their $\om=0.897$).

For small $A$, the phase transition is second order, whereas for large $A$ it becomes first order, as indicated by the sharp change in color in Fig.~\ref{phaseAMa}. It is important to note that this phase diagram is obtained using a ferromagnetic initial condition, $m_0=1$. For $\beta \gg 1$, the stable ferromagnetic region extends toward $A=1$.

However, a coexistence region of ferromagnetic and paramagnetic phases exists within the first-order regime, as shown in Fig.~\ref{phaseAMb}. Figure~\ref{phaseAMb} is obtained by computing the difference in $\langle m\rangle$ for two initial conditions, $m_0=0.99$ and $m_0=0.01$. Inside the highlighted region, both ferromagnetic and paramagnetic solutions are stable, and the observed transition boundary depends on the chosen initial condition. If either driving amplitude $A$ or temperature $T$ are varied quasistatically from the ferromagnetic phase, only the upper boundary of the coexistence region is observed. In contrast, when starting from the paramagnetic phase, no transition occurs until the lower boundary is reached. In Section~\ref{floan}, we prove that for sufficiently large $\omega$, a coexistence region always exists.\\

\begin{figure}[H]
 \centering
      \begin{subfigure}{0.49\textwidth}
         \centering
         \def\svgwidth{0.8\linewidth}        
        \includegraphics[scale = 0.37]{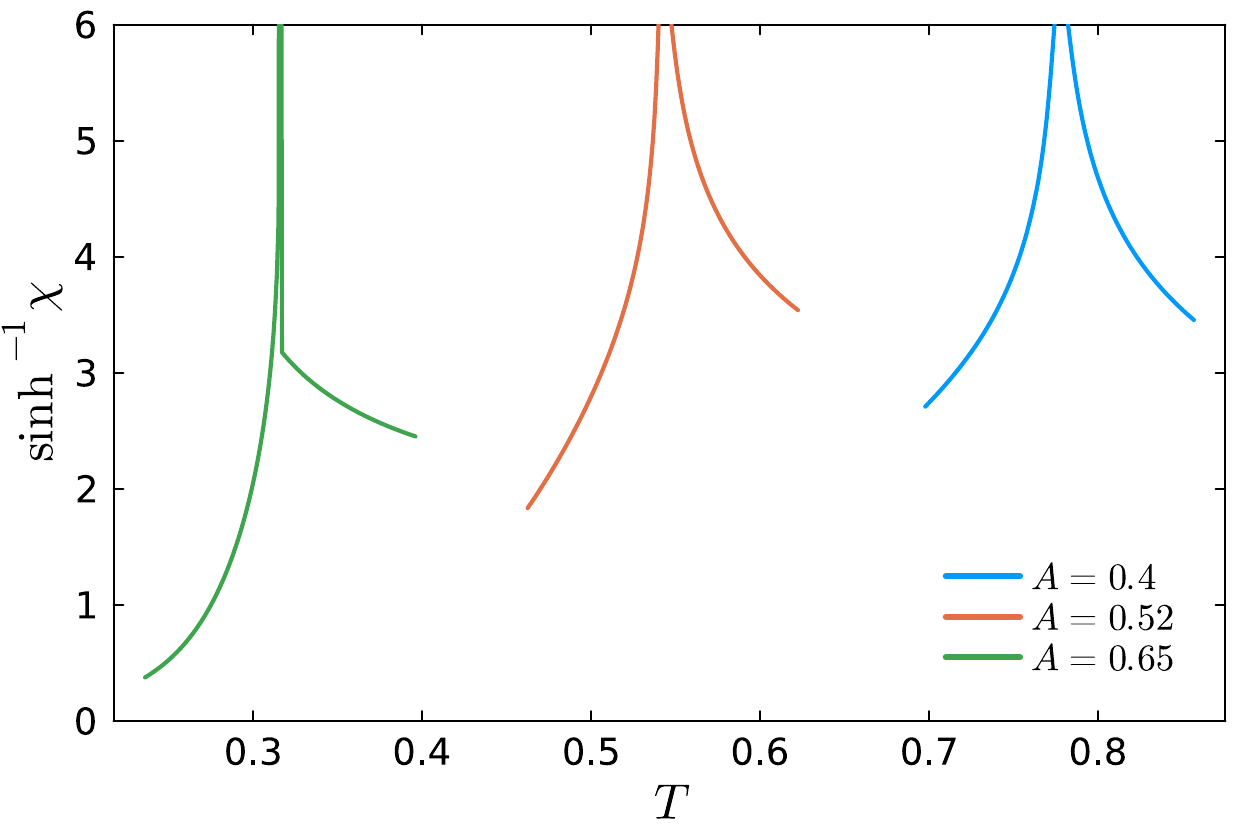}
        \caption{\small{}}
     \end{subfigure}
     \hfill
     \centering
      \begin{subfigure}{0.49\textwidth}
         \centering
         \def\svgwidth{0.8\linewidth}        
        \includegraphics[scale = 0.37]{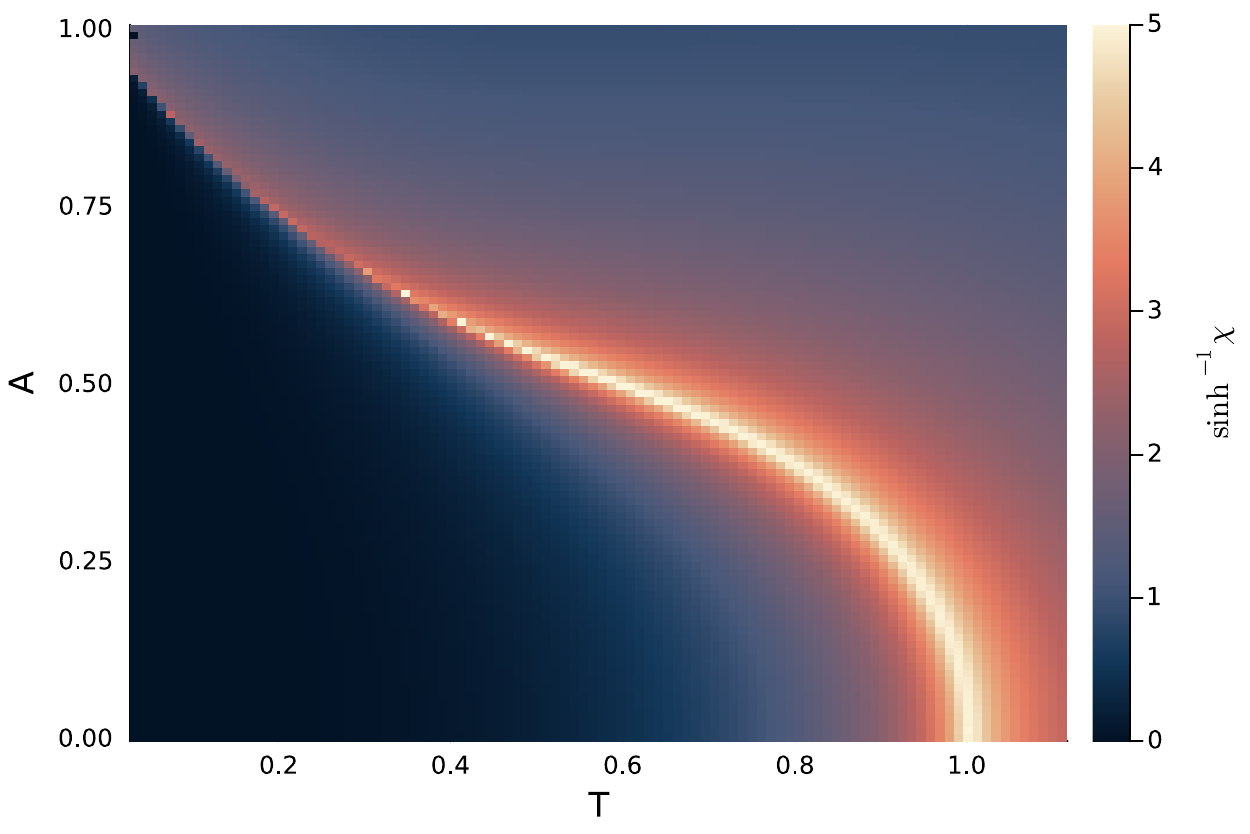}
        \caption{\small{}}
     \end{subfigure}
\caption{\small{Magnetic susceptibility at $\omega=0.9$ computed from the ferromagnetic initial condition $m_0=1$. (a) $\sinh^{-1}\chi$ as a function of temperature for several driving amplitudes $A$. (b) Phase diagram obtained from $\sinh^{-1}\chi$ in the $(A,T)$ plane.} }  \label{chiphase}
\end{figure}

It is instructive to examine the phase diagram using the susceptibility \eqref{xi}. In \fig \ref{chiphase} we plot $\sinh^{-1}\chi$ to reveal the underlying features. The phase diagram is very similar to the diagram obtained by studying $m$. 

The divergence of $\chi$ is similar to that in equilibrium. Lighter colors correspond to the larger values of $\chi$, for example, if $\sinh^{-1}\,\chi=5$, then $\chi\sim74$ which is much larger than $\sinh^{-1}\,\chi=0.5$, where $\chi\sim 0.5$.\\

\begin{figure}[H]
 \centering
      \begin{subfigure}{0.49\textwidth}
         \centering
         \def\svgwidth{0.8\linewidth}        
        \includegraphics[scale = 0.37]{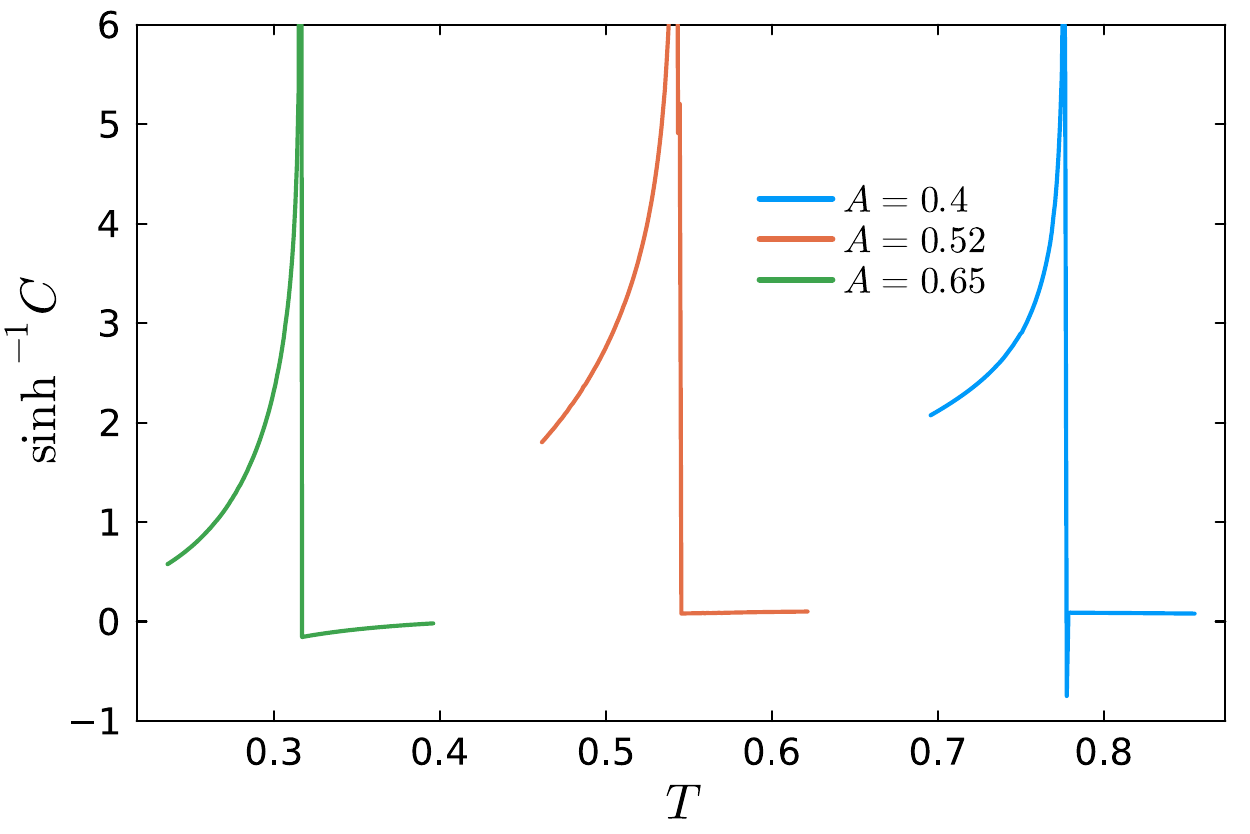}
        \caption{\small{}}
     \end{subfigure}
     \hfill
     \centering
      \begin{subfigure}{0.49\textwidth}
         \centering
         \def\svgwidth{0.8\linewidth}        
        \includegraphics[scale = 0.37]{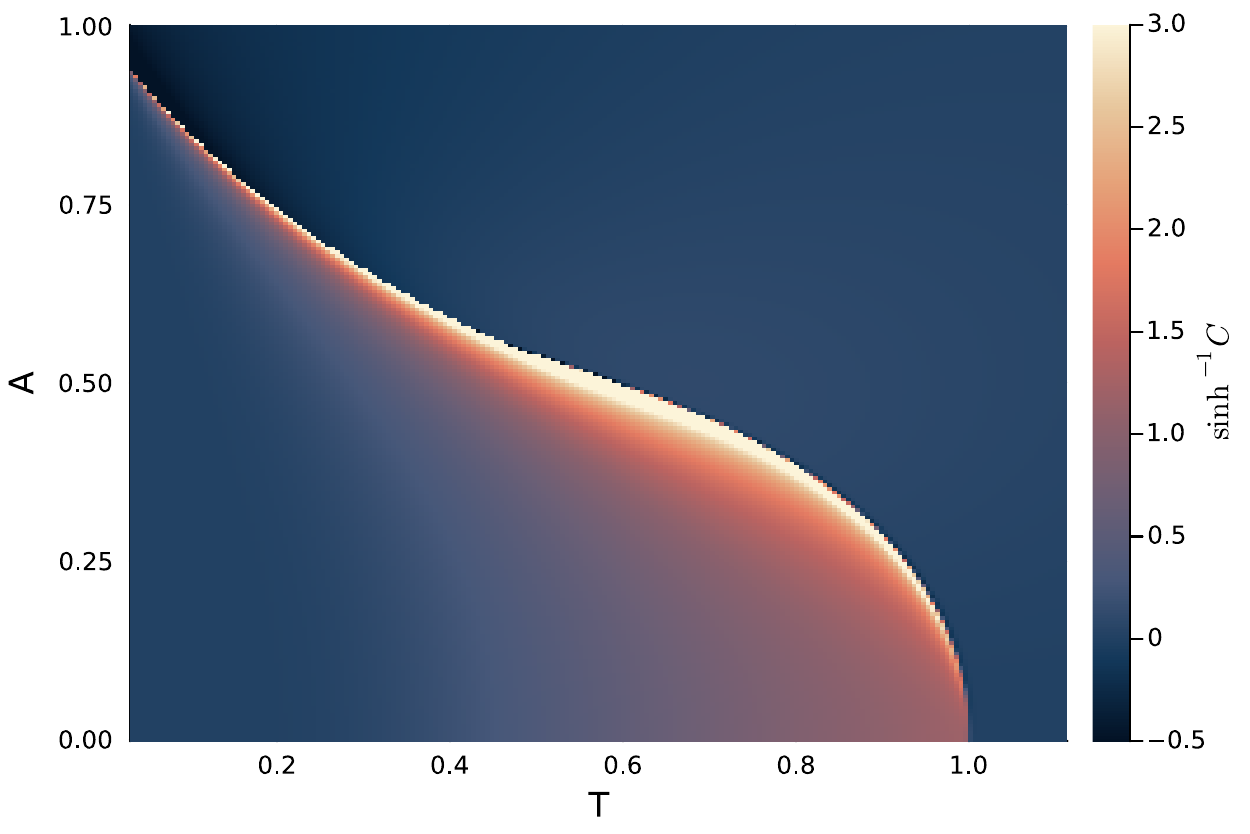}
        \caption{\small{}}
     \end{subfigure}
\caption{\small{Specific heat at $\omega=0.9$ computed from the ferromagnetic initial condition $m_0=1$. (a) $\sinh^{-1}C$ as a function of temperature for different $A$ in the first-order regime; note the regions of negative specific heat. (b) Phase diagram in the $(A,T)$ plane.
} }  \label{Cphase}
\end{figure}

\begin{figure}[H]
    \centering 
    \includegraphics[width=0.45\textwidth]{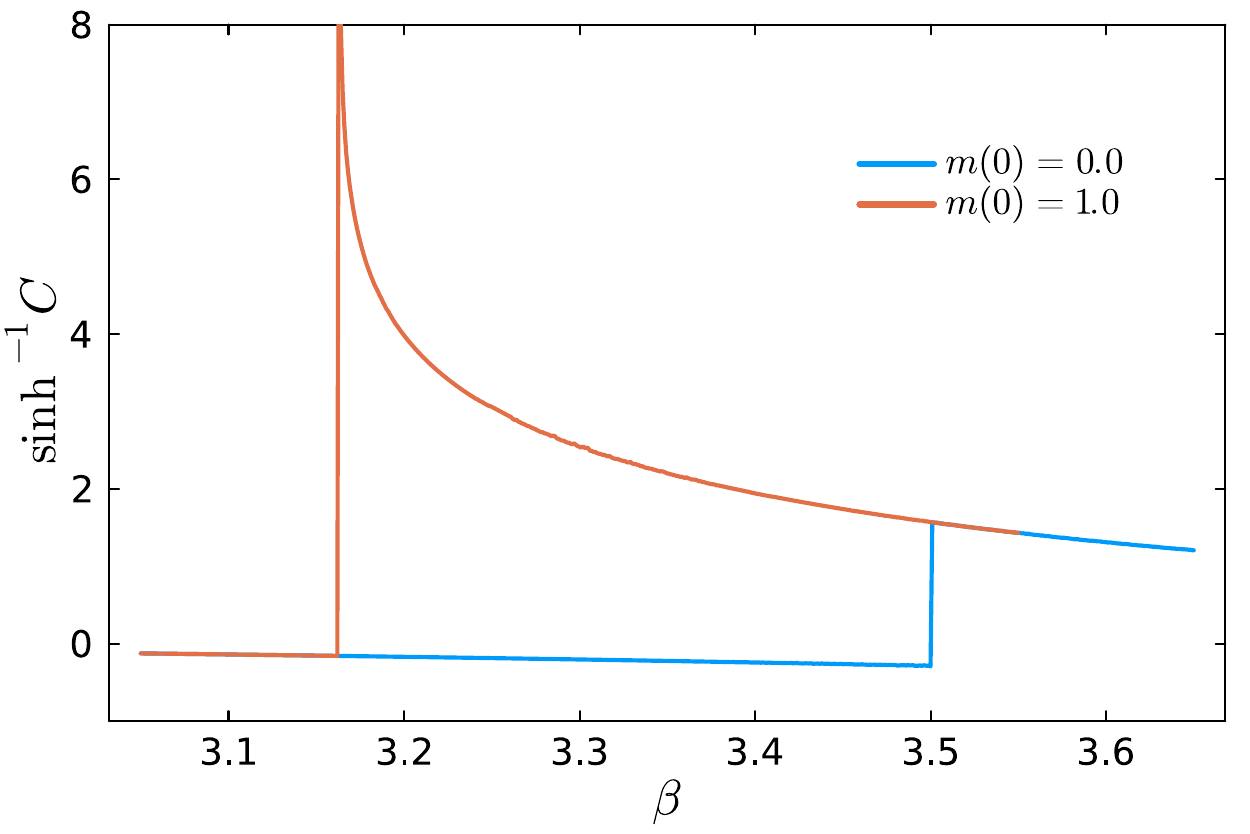} 
    \caption{First-order behavior of specific heat $C$ for $\om=0.9,A=0.65$. The two curves represent ferromagnetic ($m_0=1.0$) and paramagnetic ($m_0=0.0$) initial conditions, respectively. }
    \label{disc}
\end{figure}

For completeness, we superimpose the specific heat $C$ on the phase diagram to highlight its variation across the transition; see \fig \ref{Cphase}. 
As shown in Fig.~\ref{disc}, the specific heat depends on the initial condition, giving rise to distinct ferromagnetic ($m_0=1.0$) and paramagnetic ($m_0=0.0$) branches. The coexistence of these branches is consistent with phase coexistence and the discontinuous nature of a first-order phase transition.

\section{Coexistence of phases}\label{floan}

Recall \eqref{mag} (with $\nu=1$) for the time-evolution of the Curie--Weiss model under a time-periodic magnetic field with period $ \tau= 2\pi/\omega$.  Floquet theory provides a natural approach to the stability analysis of periodic solutions, \cite{Chicone2006}  stable and unstable dynamical phases can  be distinguished through the Floquet spectrum, allowing the identification of dynamical phase boundaries in the driven Curie--Weiss model.  With that goal, we linearize the dynamics around a periodic magnetization trajectory. The corresponding Floquet multiplier determines whether small perturbations decay or grow over one driving period.\\

\subsection{Stability criteria}
Let $m_*(t)$ denote a $ \tau$-periodic solution of \eqref{mag}.
We perturb it according to $m(t)=m_*(t)+u(t),$ where $|u(t)|\ll1.$ Linearizing \eqref{mag} yields 
 a linear ODE with periodic coefficient,
\begin{equation}\label{variational}
\dot u =a(t)u,\qquad a(t)=-1+\beta\sech^2\big( \beta\,[m_*(t)+A\cos(\omega t)]\big)
\end{equation}
where $a(t) = a(t+\tau)$. The solution is $u(t)=u(0)\exp[\int_0^t a(s)\,\id s].$ Hence, the Floquet multiplier is
\begin{equation}\label{mu}
\mu =\exp[\int_0^\tau a(t)\,\id t]
\end{equation}
and the corresponding Floquet exponent is
\begin{equation}
\lambda_F=\frac1 \tau\log\mu=-1+\frac{\beta}{ \tau}\int_0^ \tau \sech^2\big(\beta[m_*(t)+A\cos(\omega t)]\big)\,\id t
\label{floquet}
\end{equation}
The periodic solution $m_*(t)$ is linearly stable iff $\mu<1$ (equivalently $\log\mu<0$). 

\begin{figure}[H]
  \centering
  \includegraphics[width=0.5\textwidth]{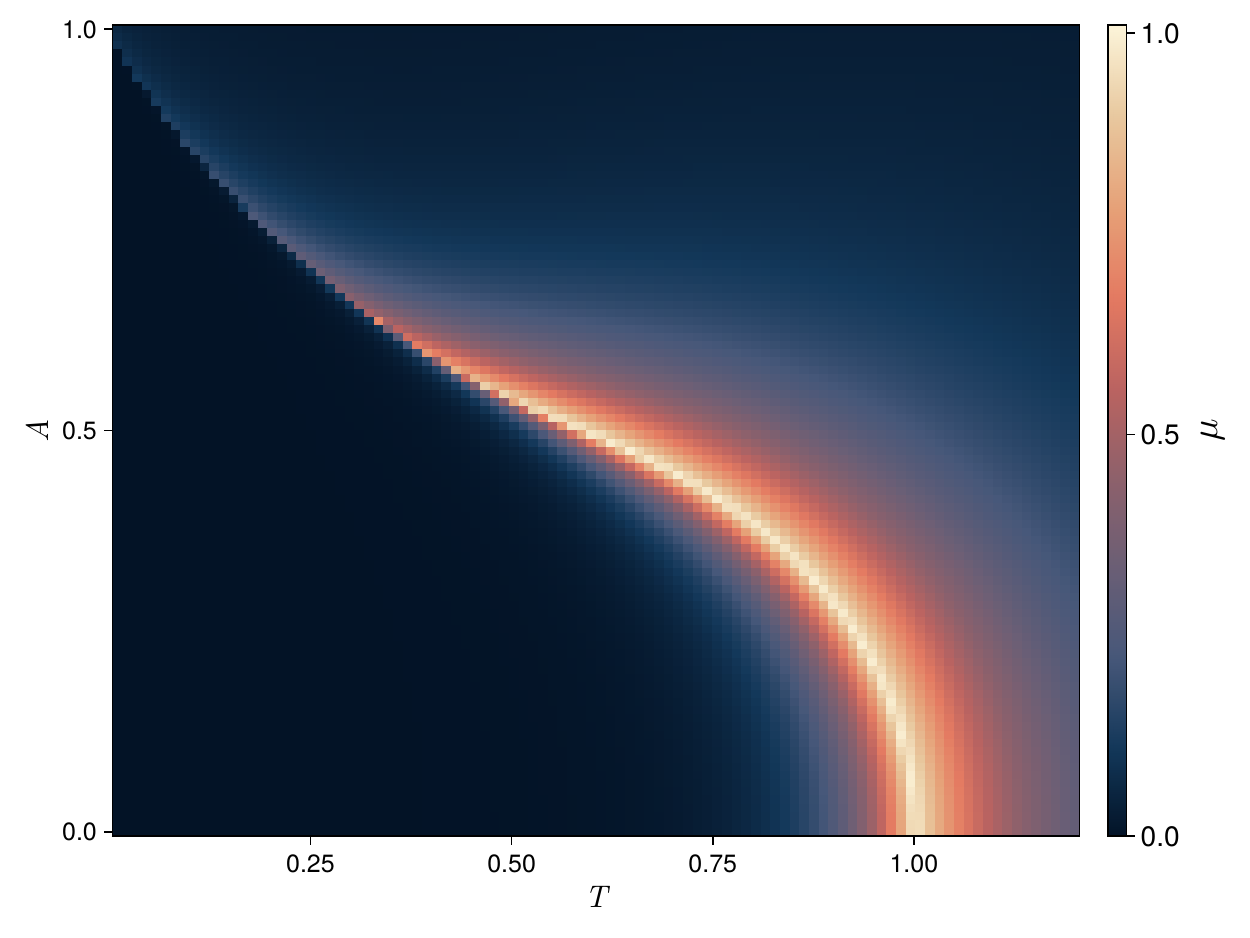}
  \caption{\small{Floquet multiplier $\mu$ in the  driving amplitude and temperature $(A,T)$ plane for $\omega =0.9$ and $m_0=1$.  The lightest colour indicates the transition region of marginal stability where  $\mu= 1$.}}
  \label{phhaseF}
\end{figure}

As a result the transition between stable and unstable behavior occurs at $\mu=1$ or $\lambda_F=0$. \\
The \textit{ferromagnetic phase} is characterized by two stable nonzero periodic solutions, $\pm m_{*} \neq 0$, corresponding to spontaneous symmetry breaking. At the transition, the Floquet multiplier reaches the critical value $\mu=1,$ where the periodic solution becomes marginally stable. This condition defines the dynamical phase boundary in the temperature and driving amplitude plane $(T,A)$; see \fig \ref{phhaseF}, which is comparable with Figs. \ref{phaseAM}--\ref{Cphase}

The \textit{paramagnetic phase} is characterized by a periodic solution with $m_{*}\approx 0$, when stable under small perturbations. Its stability is determined by the condition that the time-averaged $\beta \, \operatorname{sech}^{2}\!\big(\beta A\cos(\omega t)\big) < 1$.

\subsection{Stability of the ferromagnetic phase}
We now derive an explicit formula for $\log\mu$ at low temperature ($\beta\gg 1$) when the
driving amplitude is just below one, $A=1-\epsilon_0$ with $\epsilon_0\geq 0$.

At low temperature, $\sech^2(\beta x)$ is sharply peaked near $x=0$ and exponentially
small elsewhere.  For the ferromagnetic solution $m_*(t)\approx 1$, the argument of
$\sech^2$ in \eqref{mu} is
\[
  \beta\bigl[m_*(t)+A\cos(\omega t)\bigr]
  \approx \beta\bigl[1+(1-\epsilon_0)\cos(\omega t)\bigr]
\]
This is smallest (and hence $\sech^2$ is largest) near the times $t_0$ where
$\cos(\omega t_0)=-1$.  Expanding around such $t_0$,
%\[
 % \cos(\omega t) \approx -1 + \tfrac{\omega^2}{2}(t-t_0)^2,
%\]
%so that
\[
  1+(1-\epsilon_0)\cos(\omega t)\approx \epsilon_0 + \tfrac{(1-\epsilon_0)\omega^2}{2}(t-t_0)^2 \approx \epsilon_0 + \tfrac{\omega^2}{2}(t-t_0)^2,
\]
where in the last step we have used $\epsilon_0\ll 1$.  Using the asymptotic identity $\sech^2(x)\approx 4e^{-2x}$ for $x\gg 1$, making
\begin{equation}\label{sech_gauss}
  \sech^2\!\Bigl(\beta\bigl(\epsilon_0+\tfrac{\omega^2}{2}(t-t_0)^2\bigr)\Bigr)
  \approx 4\exp\!\Bigl(-2\beta\epsilon_0 - \beta\omega^2(t-t_0)^2\Bigr).
\end{equation}
we substitute into \eqref{mu} and we extend the Gaussian integral to $\mathbb{R}$ (valid because the integrand is negligible away from $t_0$ for large $\beta$):
\begin{align} \label{gaussian_integral1}
\int_0^\tau \beta\,\sech^2\!\Bigl(\beta\bigl(\epsilon_0+\tfrac{\omega^2}{2}(t-t_0)^2\bigr)\Bigr)\,dt
&\approx4\beta\,e^{-2\beta\epsilon_0}\int_{-\infty}^{\infty}e^{-\beta\omega^2(t-t_0)^2}\,dt\notag\\
&=4\beta\,e^{-2\beta\epsilon_0}\sqrt{\frac{\pi}{\beta\omega^2}}=\frac{4\sqrt{\pi\beta}}{\omega}\,e^{-2\beta\epsilon_0}.
\end{align}

Inserting \eqref{gaussian_integral1} into \eqref{floquet} (and noting $\tau=2\pi/\omega$):
\begin{align}\label{logmu_asymp}
  \log\mu \approx -\tau + \frac{4\sqrt{\pi\beta}}{\omega}\,e^{-2\beta\epsilon_0}= -\frac{2\pi}{\omega} + \frac{4\sqrt{\pi}}{\omega}\sqrt{\beta}\,e^{-2\beta\epsilon_0}.
\end{align}

Note that $\log\mu$ is indeed dimensionless: $\tau$ has units of time, but we have set the relaxation rate to unity ($\nu=1$), so time is measured in units of $\nu^{-1}$ and all terms are dimensionless.\\

Equation~\eqref{logmu_asymp} gives a quantitative criterion for the stability of the ferromagnetic phase at low temperature.  The first term $-2\pi/\omega$ is negative and comes from the overall relaxation over one period; it tends to stabilise $m_*$.  The second term is positive and comes from the short window near $t_0$ where the field almost cancels the magnetisation; it tends to destabilise $m_*$.  The competition between these
two terms determines the phase boundary.

\paragraph{Critical amplitude: scaling of $\epsilon_0$.}
Setting $\log\mu=0$ in \eqref{logmu_asymp} gives the critical value of $\epsilon_0$ at the
stability boundary:
\[
  \frac{2\pi}{\omega} = \frac{4\sqrt{\pi}}{\omega}\sqrt{\beta}\,e^{-2\beta\epsilon_0}
  \quad\Longrightarrow\quad
  e^{-2\beta\epsilon_0} = \frac{\sqrt{\pi}}{2}\,\beta^{-1/2}.
\]
Taking logarithms:
\begin{equation}\label{epsilon_crit}
  \epsilon_0
  = \frac{1}{2\beta}\Bigl(\log\frac{2}{\sqrt{\pi}}+\tfrac{1}{2}\log\beta\Bigr)
  \approx \frac{\log\beta}{4\beta}
  \qquad (\beta\to\infty).
\end{equation}
This means that the ferromagnetic solution remains stable even for amplitudes that are exponentially close to $A=1$, but the stability window $[0,1-\epsilon_0]$ shrinks slowly as $\beta\to\infty$.  In other words, at very low temperature, the ferromagnetic phase is \emph{robust} against large driving amplitudes, but the margin decreases logarithmically with $\beta$.

\paragraph{Instability at $A=1$.}
At $\epsilon_0=0$, equation~\eqref{logmu_asymp} gives
\[
  \log\mu \approx -\frac{2\pi}{\omega} + \frac{4\sqrt{\pi\beta}}{\omega} \to +\infty
  \quad\text{as }\beta\to\infty,
\]
so $\mu >1$ for all large $\beta$.  This proves that the ferromagnetic periodic solution \emph{loses stability} when $A=1$ in the low-temperature limit.

\paragraph{No small ferromagnetic oscillations for $A>1$.}
For $A>1$, no stable periodic solution of the form $m_0(t)=1+O(e^{-\beta})$ exists at low temperature.  To see why, suppose for contradiction that such a solution exists.  At times $t_*$ where $\cos(\omega t_*)=-1$, the argument of $\tanh$ in \eqref{mag} satisfies 
\[
  m_0(t_*)+A\cos(\omega t_*) \approx 1-A < 0,
\]
which is negative and of order unity for $A>1$.  At low temperature, $\tanh(\beta x)\approx \operatorname{sgn}(x)$, so $\tanh\!\bigl(\beta(m_0(t_*)+A\cos(\omega t_*))\bigr)\approx -1$. The evolution equation then gives $\dot{m}(t_*)+m_0(t_*)\approx -1$, hence $\dot{m}(t_*)\approx -2$. This is an $O(1)$ drift \emph{away} from $m\approx 1$, which contradicts the assumption that $m_0(t)$ remains $O(e^{-\beta})$-close to $1$ over a full period.\\

\begin{figure}[H]
 \centering
      \begin{subfigure}{0.49\textwidth}
         \centering
         \def\svgwidth{0.8\linewidth}        
        \includegraphics[scale = 0.37]{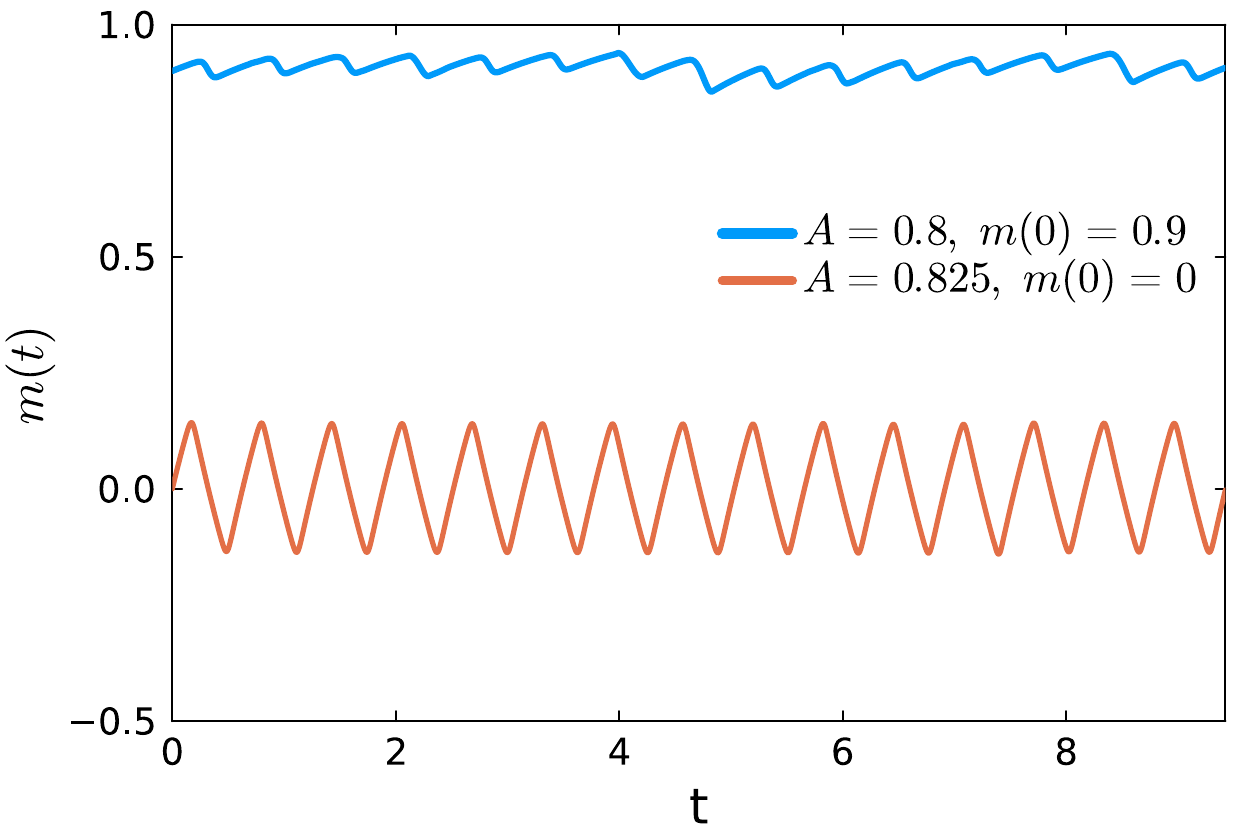}
        \caption{\small{coexistence of solutions, $\omega=10, \beta=5$}}
     \end{subfigure}
     \hfill
     \centering
      \begin{subfigure}{0.49\textwidth}
         \centering
         \def\svgwidth{0.8\linewidth}        
        \includegraphics[scale = 0.37]{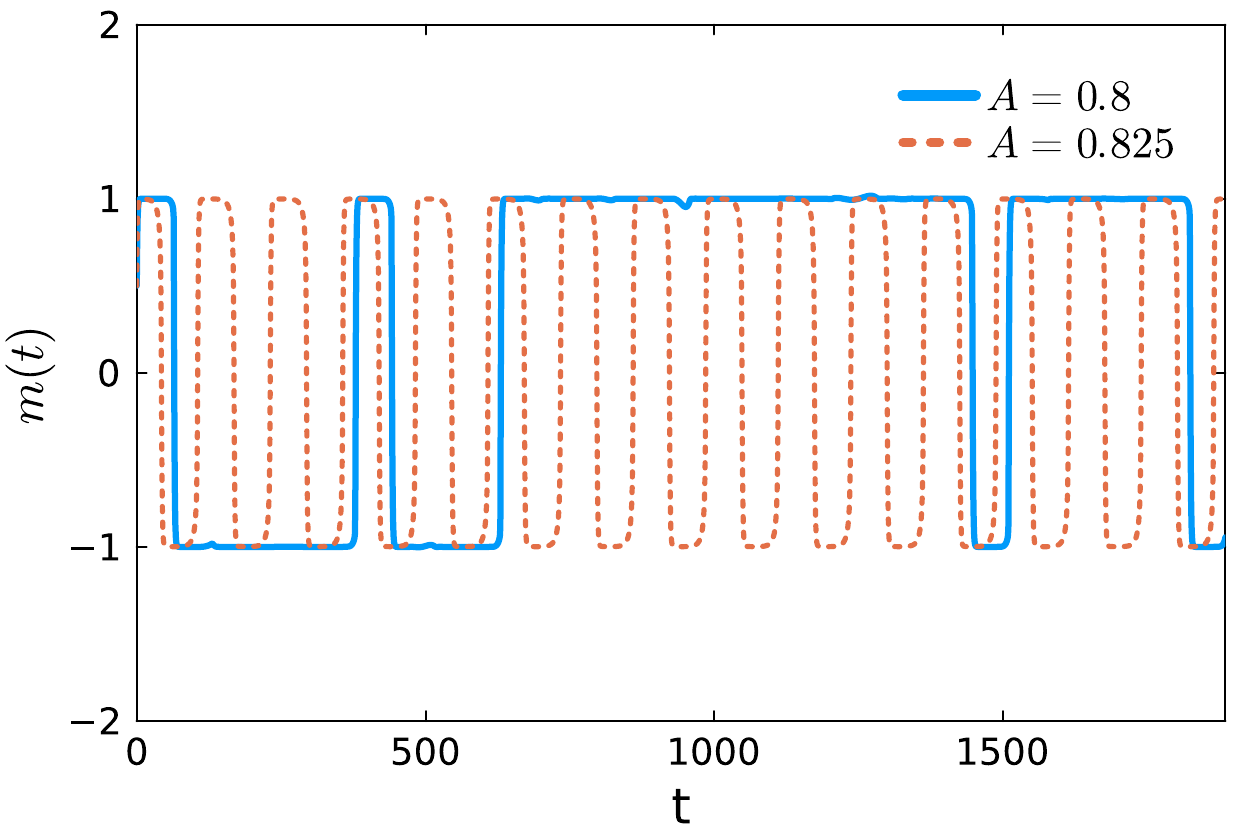}
        \caption{\small{chaotic solution, $\om=0.05, \beta=12$. }}
     \end{subfigure}
\caption{\small{Simulated trajectories. (a) For large $\om\gg1$ and $A<1$, there is coexistence of ferromagnetic and paramagnetic solution at low temperature. (b) For small $\om\ll1$ the low-temperature behavior is quite different. There is no coexistence of periodic solution. Instead, the solution will take a longer time to relax to the periodic solution as $T\to 0, A\to 1$, and is strongly affected by the initial condition. That is reminiscent of chaotic behavior. } }  \label{..}
\end{figure}

The analysis above tells us \emph{when} the ferromagnetic solution loses stability, but not what happens  afterward.  Numerical simulations suggest that for small driving frequency ($\omega\ll 1$), relaxation to any periodic solution becomes extremely slow as $\beta\to\infty$ and $A\to 1$, and the system instead exhibits irregular, switching-like dynamics that are strongly sensitive to initial conditions (see Fig.~\ref{..}b).  By contrast, for large $\omega$, the system can settle into a stable paramagnetic periodic orbit even after the
ferromagnetic one becomes unstable---a coexistence scenario we analyse next. The point $(A=1,\,\beta=\infty)$ therefore appears to act as an ``essential'' critical point of the driven Curie--Weiss model.

\subsection{Stability of the paramagnetic phase at large frequency}
\label{sec:para_stability}

The previous analysis showed that the ferromagnetic phase can lose stability for large driving amplitudes at low temperature. Here we show that for sufficiently large driving frequency $\omega$, a paramagnetic periodic solution with $m(t)\ll 0$ can be simultaneously stable.  This coexistence of two stable phases for the same control values is a  nonequilibrium phenomenon with no equilibrium analogue.\\

For a paramagnetic solution we must have a reference magnetization $m_*(t)\approx 0$ around zero  and we rewrite \eqref{floquet} as
\begin{equation}\label{mu_para}
  \log\mu 
  = \int_0^\tau \!\Bigl(-1+\beta\,\sech^2\!\bigl(\beta(m_*(t)+A\cos(\omega t))\bigr)\Bigr)\,\id t
\end{equation}
At low temperature ($\beta\gg 1$) the function $\beta\,\sech^2(\beta x)$ concentrates near the zeros of $x$, so the integral is dominated by times $t_i$ where $m_*(t_i)+A\cos(\omega t_i)=0$.
More precisely, using $\beta\,\sech^2(\beta x)\to 2\delta(x)$ as $\beta\to\infty$ (after integrating over $x$), we approximate 
\begin{equation}\label{delta}
  \beta\,\sech^2\!\bigl(\beta(m_*(t) + A\cos(\omega t))\bigr)
  \approx \sum_i \frac{2\,\delta(t-t_i)}{|\dot{m_*}(t_i)+\dot{h}(t_i)|},
\end{equation}
where $h(t)=A\cos(\omega t)$, $\dot{h}=dh/dt$, and the sum is over all zeros of  $m_*(t)+h(t)$ in one period.  The prefactor $2/|\dot{m_*}+\dot{h}|$ converts the delta-function in $x$ to a delta-function in $t$.\\
Since $m_*(t)\approx 0$ and $A=O(1)$, the condition $m_*(t)+h(t)=0$ reduces to $h(t)\approx 0$, i.e., $\cos(\omega t)=0$.  This occurs twice per period, at $t_1=\pi/(2\omega)$ and $t_2=3\pi/(2\omega)$.  At these points $\dot{h}(t_i)=-A\omega\sin(\omega t_i)$, so $|\dot{h}(t_i)|=A\omega$.\\
Substituting \eqref{delta} into \eqref{mu_para},
\begin{equation}\label{logmu_para}
  \log\mu 
  \approx -\tau + \sum_{i=1}^{2}\frac{2}{|\dot{h}(t_i)|}
  = -\frac{2\pi}{\omega} + \frac{4}{A\omega}
\end{equation}
In other words, the paramagnetic solution is stable when
\[
  -\frac{2\pi}{\omega} + \frac{4}{A\omega} < 0
  \quad\Longleftrightarrow\quad
  A > \frac{2}{\pi}.
\]
That is, for $A>2/\pi\approx 0.637$ and in the limit of large $\omega$ and large $\beta$,
the paramagnetic periodic orbit is linearly stable.

The stability of the paramagnetic phase comes from an interplay between two effects.  The term $-2\pi/\omega$ represents the natural damping of the magnetisation over one driving period; it stabilises $m_*\approx 0$.  The term $4/(A\omega)$ represents the destabilizing kicks the system receives each time the field crosses zero (twice per period), with each kick of size $2/(A\omega)$.  For large $\omega$, the kicks become small, and the damping wins --- hence the paramagnetic phase becomes stable.  The critical amplitude $A=2/\pi$ is
determined by balancing these two contributions.\\

\begin{figure}[H]
  \centering
  \includegraphics[width=0.55\textwidth]{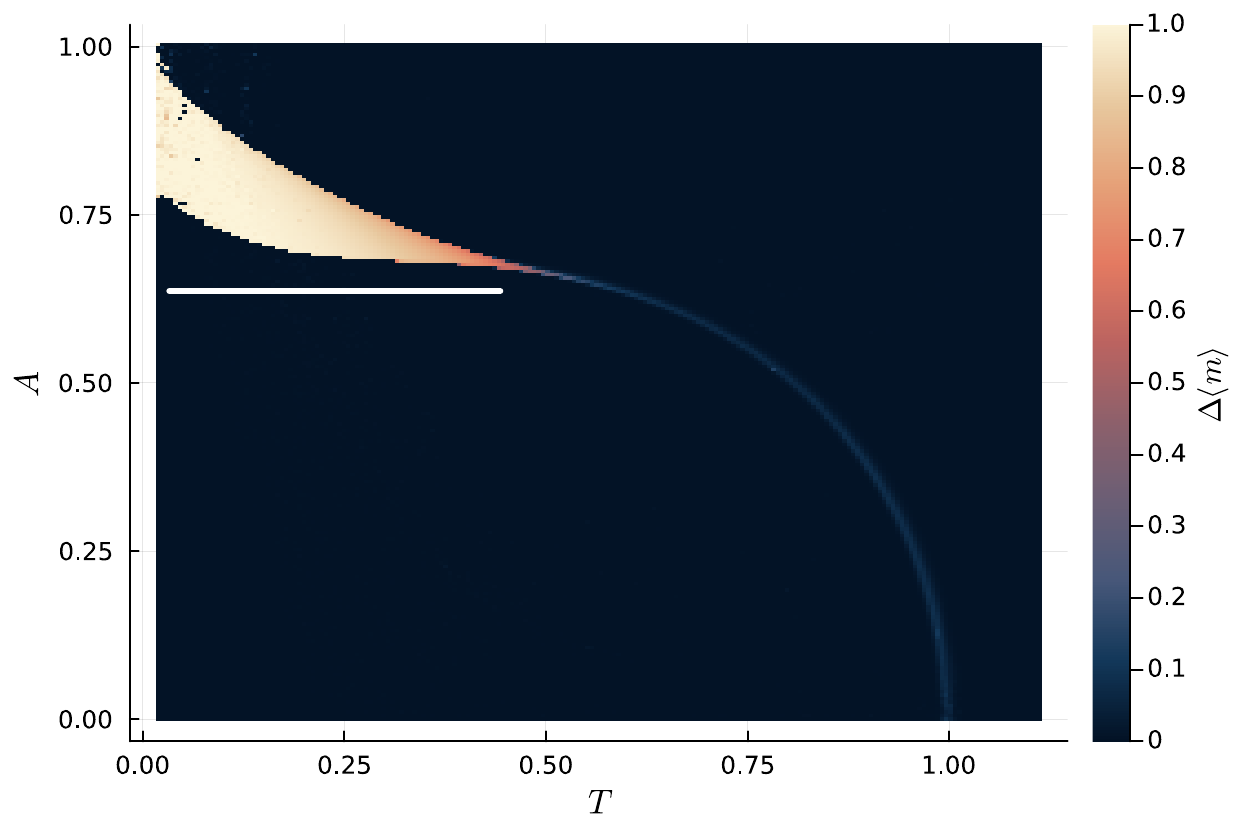}
  \caption{\small{ Coexistence region in the driving amplitude and temperature plane $(A,T)$ phase diagram for
  $\omega=2.0$, computed with two initial conditions $m_0=0$ (paramagnetic start) and $m_0=1$ (ferromagnetic start).  The white line marks the analytical paramagnetic stability boundary $A=2/\pi$.  The tricritical point sits at the edge of the coexistence region where the two phase boundaries meet.   See also~\cite{TomeOliveira1990}, Fig.~6.}}
  \label{coex1}
\end{figure}

Because the condition $A>2/\pi$ is compatible with the regime where the ferromagnetic
solution is \emph{also} stable, the two phases can
coexist for the same values of $A$, $\omega$, and $\beta$.  Fig.~\ref{coex1} shows the
coexistence region in the $(A,T)$-diagram for $\omega=2.0$: inside the coexistence
region, which phase the system selects depends on the initial condition.  The white curve
marks the paramagnetic stability boundary $A=2/\pi$.  The tricritical point at the edge of
the coexistence region is where the two boundaries meet.\\
Such coexistence, with the paramagnetic orbit stabilized by the driving, has no
equilibrium analogue and is reminiscent of the Kapitza pendulum, and also occurs in the two-temperature Curie-Weiss model of \cite{Beyen_2024}.

\section{Divergence of the specific heat }\label{divsp}

To analyze the thermal response, we introduce a small slow modulation of the inverse temperature with frequency $\omega_B\ll \omega$, where $\omega$ is the driving frequency of the magnetic field \eqref{tmod}: $\beta_t=\beta(1 +\ve \sin(\omega _B t))$.   In this regime, $m(t)=m_0(t)+\ve \,m_1(t)+O(\ve^2)$ and one may adopt a near-quasistatic ansatz for the first-order correction of the magnetization: 
\begin{equation}\label{prove1}
m_1(t) = \sigma_1(t)\sin(\omega_B t) + \omega_B \sigma_2(t)\cos(\omega_B t) + O(\omega_B^2)
\end{equation}
where $\sigma_1(t)$ and $\sigma_2(t)$ are $\tau$-periodic functions of the fast timescale $2\pi/\omega\ll 2\pi/\omega_B$. For small $\omega_B$  
\begin{equation}\label{prove2}
\dot{m}_1 =\dot{\sigma}_1\sin(\omega_B t)+\omega_B\bigl(\sigma_1+\dot{\sigma}_2\bigr)\cos(\omega_B t) +O(\omega_B^2)
\end{equation}
On the other hand, the linearized evolution equation \eqref{mag} (similar to \eqref{m1m0dot} but for small perturbation $\ve$) reads
\begin{align}\label{prove3}
\dot{m}_1 +J(t)m_1 = \beta F(t)\sin(\omega_B t) +O(\ve^2)
\end{align}
where we have defined
\begin{equation}\label{jt}
J(t) = 1-\beta \sech^2(\beta(m_0+h_t)),
\qquad 
F(t) = \sech^2(\beta(m_0+h_t))(m_0+h_t).
\end{equation}
Note that $J(t)=-a(t) $ of  in \eqref{mu} if $m_0(t)=m_*(t)$, where we use here an arbitrary periodic solution $m_0(t)$.
Substituting \eqref{prove1} and \eqref{prove2} into \eqref{prove3} gives
\begin{align}
\dot{\sigma}_1 +J(t) \sigma_1=\beta F(t), \qquad \dot{\sigma}_2 +J(t)\sigma_2 =-\sigma_1
\label{Beq}
\end{align}
Putting
\begin{equation}
  \mu (t)= \exp\, [-\int_0^t J(s)\,\id s],   \qquad   \mu = \mu(\tau)=\exp[-\int_0^\tau J(s)\,\id s]
\end{equation}
the solution of \eqref{Beq} is
\begin{align}
  \sigma_1(t)
  &= \frac{\beta\,\mu(t)\mu}{1-\mu}
    \int_t^{t+\tau}\frac{F(s)}{\mu(s)}\,\id s,
    \label{sigma1}\\[4pt] 
  \sigma_2(t) &=-\frac{\mu(t)\mu}{1-\mu}
    \int_t^{t+\tau}\frac{\sigma_1(s)}{\mu(s)}\,\id s.
    \label{sigma2}
\end{align}
Substituting the Ansatz \eqref{prove1} in \eqref{p1}, discarding oscillatory terms that vanish upon integration over one period, and keeping only the $\cos^2(\omega_B t)$ contributions, one finds
\begin{align}
C &= \frac{\beta}{\pi}\, \frac{2\pi}{\omega_B}\,\frac{\omega_B}{2}
\big\langle \dot{m}_0\sigma_2+(m_0(t)+h(t))(\sigma_1+\dot{\sigma}_2)\big\rangle \\
&=\beta\big\langle \dot{m}_0\sigma_2+(m_0(t)+h(t))(\sigma_1+\dot{\sigma}_2)\big\rangle
\end{align}
where $\langle\cdot\rangle$ denotes an average over the fast driving period $\tau=2\pi/\omega$.  
By subtracting a total derivative, this can be recast as
\begin{equation}
C = \beta\,\big\langle \sigma_1(t)(m_0(t)+h(t))\big\rangle- \beta\big\langle \dot{h}(t)\sigma_2(t)\big\rangle
\label{Cformula}
\end{equation}

Eq.~\eqref{Cformula} shows that the boundedness of the specific heat is entirely controlled by the boundedness of the auxiliary functions $\sigma_1(t)$ and $\sigma_2(t)$.   
From \eqref{sigma1} and \eqref{sigma2} we see that if $\mu<1$, both $\sigma_1$ and $\sigma_2$ remain finite, so $C$ is bounded and no divergence occurs. \\

\begin{figure}[H]
 \centering
      \begin{subfigure}{0.49\textwidth}
         \centering
         \def\svgwidth{0.8\linewidth}        
        \includegraphics[scale = 0.37]{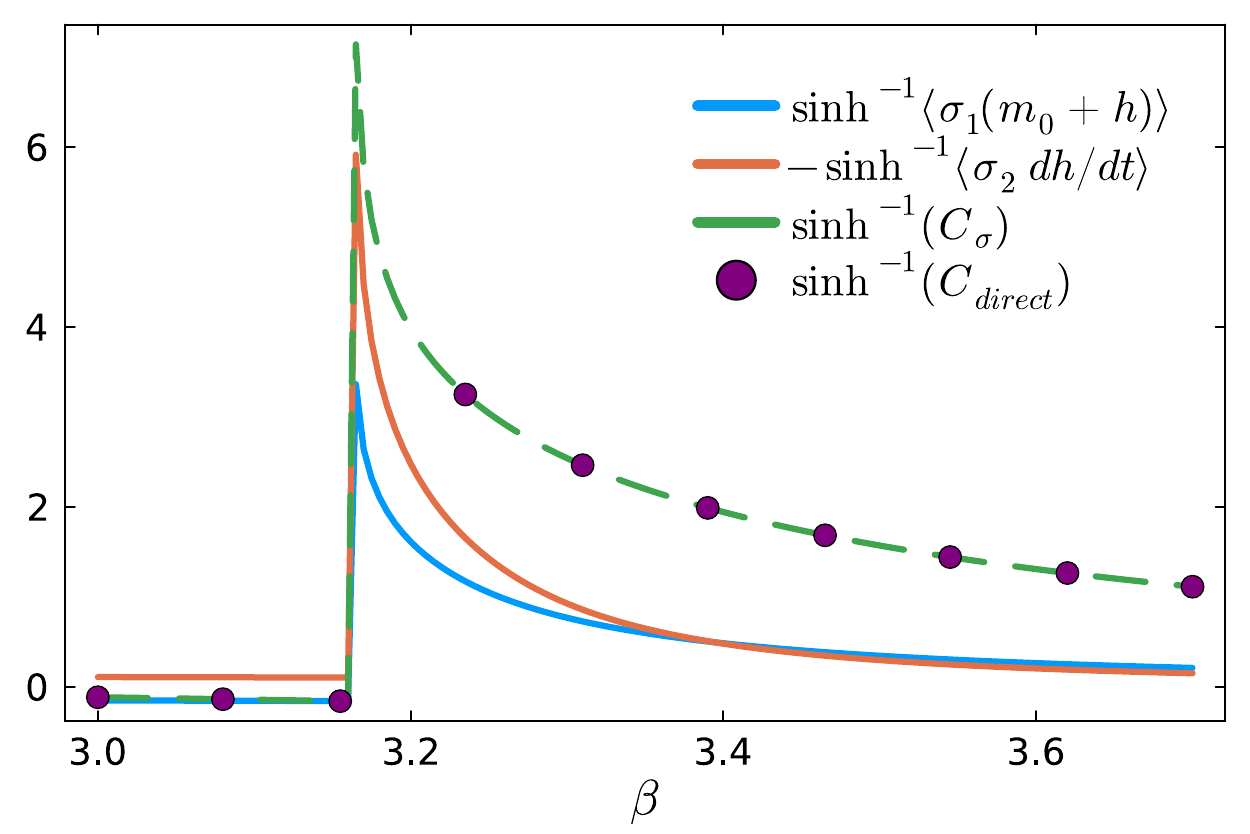}
        \caption{\small{$\omega =0.9$ and $A=0.65$}} \label{compa}
     \end{subfigure}
     \hfill
     \centering
      \begin{subfigure}{0.49\textwidth}
         \centering
         \def\svgwidth{0.8\linewidth}        
        \includegraphics[scale = 0.37]{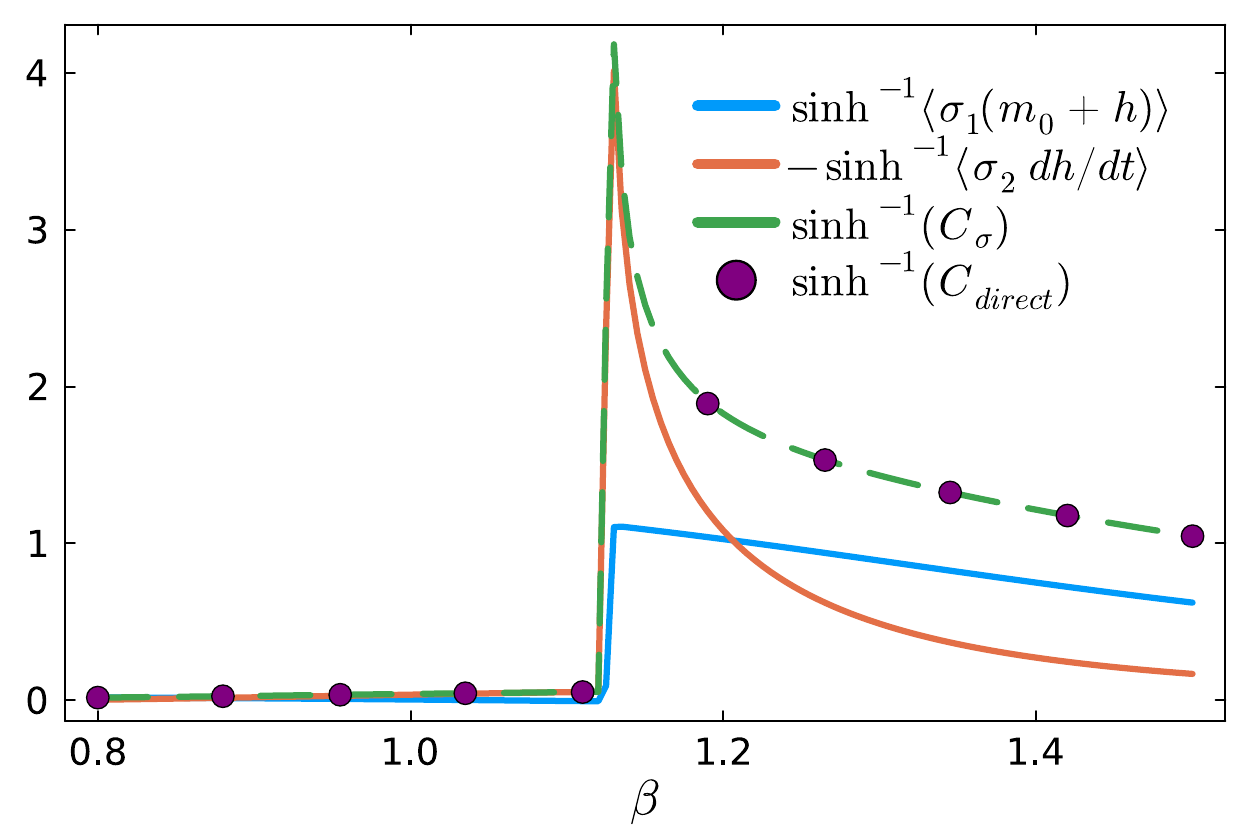}
        \caption{\small{$\omega =0.9$ and $A=0.3$ }} \label{compb}
     \end{subfigure}
\caption{\small{Contributions to the heat-capacity divergence in the regime of (a) first-order, and (b) second order transition  for $\omega = 0.9$. $C_{\text{direct}}$ denotes the heat capacity obtained from \eqref{hc}, and $C_\sigma$ is the heat capacity obtained from \eqref{Cformula}.
} }  \label{comp}
\end{figure}

Eq.~\eqref{Cformula} provides an efficient way to compute the heat capacity while separating the response into an in-phase contribution, $\sigma_1(m_0+h)$, and a dissipative contribution, $-\sigma_2\dot{h}$. At the same time, that decomposition helps identify the origin of the observed divergences.\\
Fig.~\ref{comp} shows these two contributions separately. In the regime with a first-order transition ($\omega=0.9$, $A=0.65$; Fig.~\ref{compa}), both contributions increase strongly as the transition is approached from the ferromagnetic phase, although they appear to do so with different rates. No analogous divergence is observed when approaching the transition from the paramagnetic phase.\\
In contrast, in the regime with a second-order transition ($\omega=0.9$, $A=0.3$; Fig.~\ref{compb}), the in-phase contribution remains finite at the critical point, whereas the dissipative contribution diverges. Within numerical accuracy, the divergence of the heat capacity is therefore entirely attributable to the dissipative term. Since this contribution is absent in the equilibrium limit ($A=0$, for which $\dot{h}=0$), the corresponding nonequilibrium divergence disappears in the absence of periodic driving.

\begin{figure}[H]
 \centering
      \begin{subfigure}{0.49\textwidth}
         \centering
         \def\svgwidth{0.8\linewidth}        
        \includegraphics[scale = 0.37]{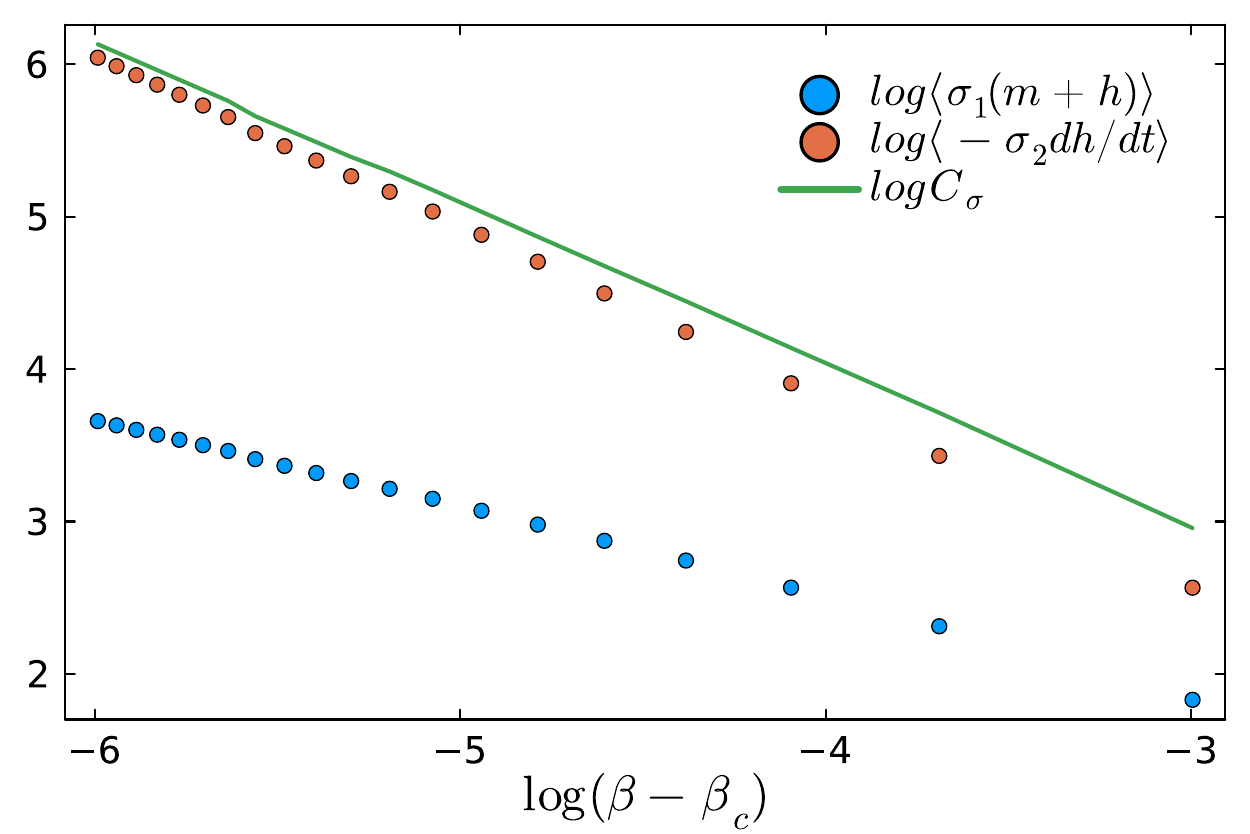}
        \caption{\small{$\omega =0.9$ and $A=0.65$}} \label{Fsopta}
     \end{subfigure}
     \hfill
     \centering
      \begin{subfigure}{0.49\textwidth}
         \centering
         \def\svgwidth{0.8\linewidth}        
        \includegraphics[scale = 0.37]{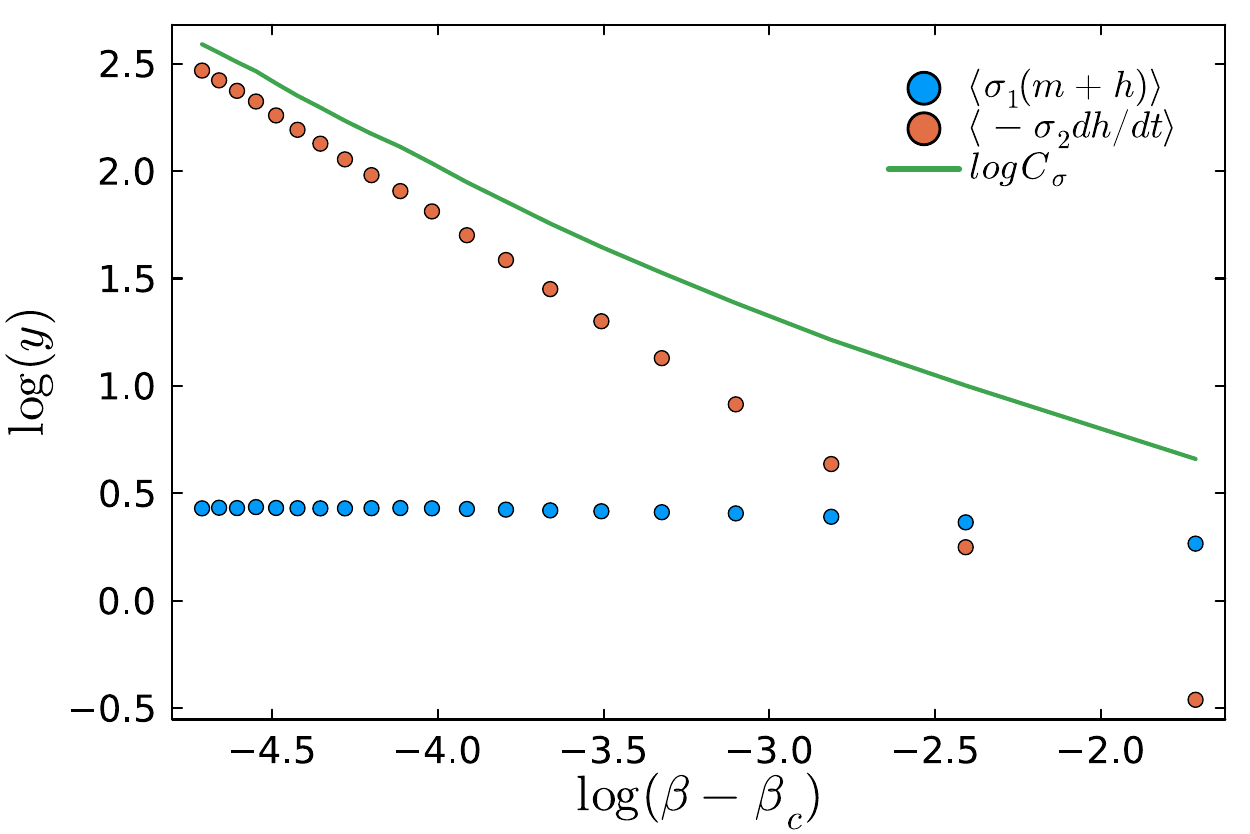}
        \caption{\small{$\omega =0.9$ and $A=0.3$ }} \label{Fsoptb}
     \end{subfigure}
\caption{\small{Logarithmic plot for the heat capacity and its components in  \eqref{Cformula}, for the (a) first-order, and  (b)  second-order region of phase transition. $C_\sigma$ is the heat capacity obtained from \eqref{Cformula}.} }  \label{Fsopt}
\end{figure}

In \fig~\ref{Fsopt}, as result of simulations, the critical exponent of the heat capacity and its components are studied. In the region of the first-order phase transition \fig \ref{Fsopta}, $\sigma_1$ and $\sigma_2$ have an exponent $ -1/2$ and $-1$, respectively, and $C$ has exponent  $-1$. In the region of the second-order phase transition \fig \ref{Fsoptb}, $\sigma_1$ and $\sigma_2$ have an exponent of $0$ and $-1$, respectively, and $C$ has an exponent of $-1$.

\section{Divergence of the susceptibility}\label{divmag}
In Section \ref{ms} the DC-susceptibility is studied, where   $\chi =\frac{1}{\tau}\lim_{s\to \infty} \int_s^{s +\tau} m_1(t)\,\id t  $ for  $m(t) = m_{0}(t) + \delta m_{1}(t)$. 
Rewriting \eqref{m1m0dot1},
\begin{align}
\dot{m}_{1} + J(t)m_{1} = g(t),
\end{align}
where $J(t)=1-\beta\,\sech^2\,\big(\beta(m_0(t)+A \cos (\omega t))\big)$ as in \eqref{jt}, and $g(t)=\beta\,\sech^2\,\big(\beta(m_0(t)+A \cos (\omega t))$.  Therefore,
\[
m_{1}(t) = \frac{1}{\mu(t)}\left( \int_{0}^{t}\id s \, \mu(s)g(s) + m_{1}(0) \right)
\]
where again \(\mu(t) = \exp( - \int_{0}^{t}J(s)ds)\) and   $\mu=\mu(\tau)$. Applying the time-periodic condition $m_{1}(0) = m_{1}(\tau)$ yields
\[
m_{1}(0) = \frac{\mu}{1 - \mu}\int_{0}^{\tau}\id s \, \mu(s)\,g(s)\]
and therefore,
\begin{align}
    \chi =\frac{1}{\tau} \frac{\mu}{1 - \mu}\int_{0}^{\tau}\id s \, \mu(s)g(s)\int_0^\tau\frac{1}{\mu(t)}\, \id t + \frac{1}{\tau}\int_{0}^{\tau}dt\frac{1}{\mu(t)}\left( \int_{0}^{t}\id s \, \mu(s)g(s) \right).
\end{align}
The second term remains bounded, and the divergence is again controlled by the way $\mu\to 1$, which determines how the relaxation time diverges.

\section{Conclusion}
Curie–Weiss models under an oscillating magnetic field represent mean-field models that are driven out of equilibrium. The competition between the relaxation time and the driving period leads to a dynamic phase transition. This transition separates a dynamically ordered phase, with nonzero period-averaged magnetization, from a dynamically disordered phase.  We have added the analysis of the specific heat and completed the analysis of the phase diagram.  It shows interesting features in the coexistence of ferromagnetic and paramagnetic phases (at large enough driving amplitude), and the transition from a second order to a first order regime.  The critical point is identifiable and we have characterized the divergence of specific heat and magnetic susceptibility.  Floquet analysis relates those to the critical slowing down in the system.\\

\noindent {\bf Acknowledgment:}  FK is supported by the Research Foundation–Flanders (FWO) postdoctoral fellowship 1232926N.

\bibliographystyle{unsrt}  %ieeetr
\bibliography{chr}
\onecolumngrid
\newpage
\appendix
\section{Proof of susceptibility criticality at small driving}\label{suscrit}
We start with the asymptotic form of the DC-susceptibility in the high-temperature regime for weak periodic driving.\\
Where $h(t)=\delta+A\cos\omega t,$ and write the magnetization as $m(t)=m_0(t)+\delta\,m_1(t)+O(\delta^2);$ where $m_0(t)$ denotes the periodic solution at $\delta=0$. Recall that the susceptibility is the period-averaged linear response
\[
\chi\langle m_1\rangle_\tau=\frac{\omega}{2\pi}\int_0^{2\pi/\omega}m_1(t)\,dt.
\]

The linearized equation for $m_1$ is
\begin{equation}
\dot m_1+p(t)m_1=q(t),\label{m1eq_appendix}
\end{equation}
with
\[
q(t)=\beta\sech^2\!\bigl(\beta(m_0(t)+A\cos\omega t)\bigr),\qquad p(t)=1-q(t).
\]
For $\beta\ll 1$, the unperturbed magnetization remains small for weak driving. %Expanding $\tanh x=x-\frac{x^3}{3}+O(x^5),$ t
The equation for $m_0$ becomes
\[
\dot m_0=-(1-\beta)m_0+\beta A\cos\omega t+O(A^3).
\]
With the reduced distance $b=\beta^{-1}-1$ from (equilibrium) criticality. we have $\dot m_0+\beta b\,m_0=\beta A\cos\omega t.$ 
Hence, the unique periodic steady-state solution is
\begin{equation}
m_0(t)=\frac{\beta A}     {(\beta b)^2+\omega^2}\Bigl(\beta b\cos\omega t+\omega\sin\omega t\Bigr)  \label{m0solution}
\end{equation}
Its mean over a period vanishes, $\langle m_0\rangle_\tau=0,$  while
\begin{equation}
\langle m_0^2\rangle_\tau=\frac{\beta^2A^2}     {2\bigl((\beta b)^2+\omega^2\bigr)}
\label{m0variance}
\end{equation} 
To leading order in $A$, we expand $\sech^2 x=1-x^2+O(x^4),$ giving
\begin{eqnarray*}
q(t)&=&\beta-\beta^3\bigl(m_0(t)+A\cos\omega t\bigr)^2+O(A^4),\\
p(t)&=&\beta b+\beta^3\bigl(m_0(t)+A\cos\omega t\bigr)^2+O(A^4).
\end{eqnarray*}
Averaging over one period gives
\begin{equation}
\langle p\rangle_\tau=\beta b+\beta^3\Bigl\langle\bigl(m_0+A\cos\omega t\bigr)^2\Bigr\rangle_\tau+O(A^4)
\label{paverage}
\end{equation}
Substituting \eqref{m0solution} and evaluating the average yields
\begin{align*}
    \Bigl\langle\bigl(m_0+A\cos\omega t\bigr)^2\Bigr\rangle_\tau&=\Bigl\langle m_0^2\Bigr\rangle_\tau+A^2\Bigl\langle \cos^2\omega t\Bigr\rangle_\tau +2A \Bigl\langle m_0.\cos\omega t\Bigr\rangle_\tau\\
    &= \frac{\beta^2A^2}     {2\bigl((\beta b)^2+\omega^2\bigr)}+ \frac{A^2}{2}
+\frac{\beta^2 A^2 b}{(\beta ^2 b^2 +\omega^2)}
\end{align*}

Next we move near the critical point $\beta\simeq 1$ and $\beta b\simeq b$, so $\beta^2 b^2+\omega^2\simeq b^2+\omega^2$. The first term $A^2/2$ is non-singular, while the second term diverges as $b\to 0$. Retaining only the singular contribution:
\begin{equation}\label{avg_square_singular}
    \Bigl\langle\bigl(m_0+A\cos\omega t\bigr)^2\Bigr\rangle_\tau    \approx \frac{A^2}{2} + \frac{A^2\!\left(\frac{1}{2}+b\right)}{b^2+\omega^2}\approx \frac{A^2/2}{\omega^2+b^2},
\end{equation}
where in the last step we used $\frac{1}{2}+b\approx\frac{1}{2}$ near the critical point ($b\to 0$) and dropped the non-singular $A^2/2$ since it generates only an $O(A^2\beta^3)$ correction to $\langle p\rangle_\tau$ which is subleading compared to the singular term. Equation~\eqref{paverage} then becomes
\begin{equation}\label{pavg_final}
    \langle p\rangle_\tau  \approx b + \frac{\beta^3 A^2/2}{\omega^2+b^2}    \approx b + \frac{A^2/2}{\omega^2+b^2}
\end{equation}
where in the last step we set $\beta\simeq 1$. To obtain $\chi$ from $\langle p\rangle_\tau$, we average 
\eqref{m1eq_appendix} over one period. Since $m_1(t)$ is periodic, $\langle\dot m_1\rangle_\tau=0$, and averaging gives
\begin{equation}\label{avgm1}
    \langle p\rangle_\tau\,\langle m_1\rangle_\tau 
    + \langle\tilde p\,\tilde m_1\rangle_\tau 
    = \langle q\rangle_\tau
\end{equation}
where $\tilde p = p - \langle p\rangle_\tau$ and $\tilde m_1 = m_1 - \langle m_1\rangle_\tau$ are the 
oscillating parts. Since $\tilde p = O(A^2)$ and $\tilde m_1 = O(A^2)$, the cross-term $\langle\tilde p\,\tilde m_1\rangle_\tau = O(A^4)$ is negligible. Using $p(t)+q(t)=1$ for all $t$, so $\langle p\rangle_\tau + \langle q\rangle_\tau = 1$, \eqref{avgm1} becomes
\begin{equation}
    \chi = \langle m_1\rangle_\tau 
         = \frac{\langle q\rangle_\tau}{\langle p\rangle_\tau}
         = \frac{1 - \langle p\rangle_\tau}{\langle p\rangle_\tau}
         = \frac{1}{\langle p\rangle_\tau} - 1
\end{equation}
Near the critical point $\chi\gg 1$, so $1/\langle p\rangle_\tau \gg 1$, and the $-1$ correction is negligible compared to $1/\langle p\rangle_\tau$. Therefore
\begin{equation}
    \chi \approx \frac{1}{\langle p\rangle_\tau}.
\end{equation}
Substituting \eqref{pavg_final}:
\begin{equation}\label{chiinversefinal}
    \chi^{-1} \approx b + \frac{A^2/2}{\omega^2+b^2},
\end{equation}
or equivalently
\begin{equation}\label{chiHTapprox}
    \chi \approx \frac{1}{\,b+\dfrac{A^2/2}{\omega^2+b^2}\,}.
\end{equation}
which is formula \eqref{highfit}.
Equation~\eqref{chiHTapprox} shows that the periodic drive produces a positive shift of the effective distance from 
criticality,
\begin{equation}
    b_{\text{eff}} = b + \frac{A^2/2}{\omega^2+b^2},
\end{equation}
which, as explained above  \eqref{effb}, suppresses the divergence of the susceptibility away from the equilibrium critical point.\\

Finally, we derive the subcritical temperature susceptibility formula \eqref{lowfit} by expanding systematically around the ferromagnetic fixed point $m_s > 0$, defined by $m_s = \tanh(\beta m_s)$, for small driving amplitude $A$.
We introduce the shorthand notation
\begin{equation}
g_1 = \beta\,\mathrm{sech}^2(\beta m_s), \qquad g_2 = \beta^2\tanh''(\beta m_s), \qquad g_3 = \beta^3\tanh'''(\beta m_s),
\end{equation}
and $d = 1 - g_1 > 0$, which is positive in the low-temperature phase since $g_1 < 1$ for $\beta > \beta_c = 1$.

We write $m_0(t) = m_s + A\,u_1(t) + A^2\,u_2(t) + O(A^3)$ and expand the zeroth-order equation. At $O(A^0)$ the fixed-point equation $m_s = \tanh(\beta m_s)$ is satisfied identically.
At order $O(A)$, linearizing gives
\begin{equation}
\dot{u}_1 = -d\,u_1 + g_1\cos\omega t
\end{equation}
whose asymptotic periodic solution %, found by the ansatz $u_1(t) = P\cos\omega t + Q\sin\omega t$, 
is
\begin{equation}
u_1(t) = \frac{g_1}{d^2+\omega^2}\bigl(d\cos\omega t + \omega\sin\omega t\bigr)
\end{equation}
It is convenient to write $u_1(t) + \cos\omega t = \alpha\cos\omega t - \gamma\sin\omega t$, where
\begin{equation}
\alpha = 1 + \frac{g_1 d}{d^2+\omega^2} = \frac{d^2+\omega^2 + g_1 d}{d^2+\omega^2}, \qquad \gamma = - \frac{g_1\omega}{d^2+\omega^2}
\end{equation}
One computes
\begin{equation}\label{algam}
\alpha^2 + \gamma^2 = \frac{(d^2+\omega^2+g_1 d)^2+g_1^2 \omega^2}{(d^2+\omega^2)^2}
\end{equation}
At order $O(A^2)$, the DC-component of $u_2$ satisfies $0 = -d\langle u_2\rangle + \tfrac{g_2}{4}(\alpha^2+\gamma^2)$, giving
\begin{equation}
\langle u_2\rangle = \frac{g_2(\alpha^2+\gamma^2)}{4d}
\end{equation}
We write $m_1(t) = m_1^{(0)} + A\,m_1^{(1)} + A^2\,m_1^{(2)} + \cdots$ and expand the first-order equation.
At $O(A^0)$, the asymptotic constant solution is
\begin{equation}
m_1^{(0)} = \frac{g_1}{d}
\end{equation}
At $O(A)$, using $m_1^{(0)} + 1 = 1/d$, the equation reads
\begin{equation}
\dot{m}_1^{(1)} = -d\,m_1^{(1)} + \frac{g_2}{d}\bigl(\alpha\cos\omega t - \gamma\sin\omega t\bigr)
\end{equation}
which is purely oscillatory, so $\langle m_1^{(1)}\rangle = 0$.
At $O(A^2)$, taking the DC part of the equation for $m_1^{(2)}$ and using the asymptotic solution $m_1^{(1)}(t) = B\cos\omega t + C\sin\omega t$ with $B = g_2(d \alpha +\omega\gamma)/d(d ^2+\omega^2)$ and $C = g_2(d \gamma -\omega \alpha)/d(d^2+\omega^2)$, the DC-balance gives
\begin{equation}
d\langle m_1^{(2)}\rangle=(\alpha^2+\gamma^2)\left(\frac{g_3}{4d}+\frac{g_2^2}{4d^2}+\frac{g_2^2}{2(d^2+\omega^2)}\right)+\frac{\,g_2^2\,\omega\,\alpha\gamma}{d(d^2+\omega^2)}
\end{equation}
The DC-susceptibility is
\begin{equation}
\chi = \langle m_1\rangle = \frac{g_1}{d} + A^2\langle m_1^{(2)}\rangle + O(A^4)
\end{equation}
Motivated by the perturbative expansion, we resume the susceptibility using 
\begin{equation}
\chi = -g_1 + \frac{g_1(1+d)}{d - \Sigma}
\end{equation}
which is the  Pad\'e approximant reproducing $g_1/d$ at $A=0$ (since $-g_1 + g_1(1+d)/d = g_1/d$) and the correct $O(A^2)$ coefficient upon expanding $1/(d-\Sigma) \approx (1/d)(1 + \Sigma/d)$.
Here $\Sigma$ encodes the $O(A^2)$ renormalization of the relaxation rate $d$ by the drive,
\begin{equation}
\Sigma = \frac{A^2 d}{g_1(1+d)}\left[\frac{\alpha^2+\gamma^2}{4}\left(\frac{g_3}{d}+\frac{g_2^2}{d^2}+\frac{2g_2^2}{d^2+\omega^2}\right)+\frac{g_2^2\,\omega\,\alpha\gamma}{d(d^2+\omega^2)}\right]
\end{equation}
Here the $\alpha^2+\gamma^2$ terms collect contributions from the cubic nonlinearity of  $\tanh$ (through $g_3$) and from the second-order parametric feedback of the drive (through $g_2^2$), while the cross term $\omega\alpha\gamma$ is odd in $\omega$ and vanishes in the static limit $\omega\to 0$. The frequency-dependent amplitudes $\alpha^2+\gamma^2$ and $\alpha\gamma$, given in~\eqref{algam}, are determined by the first-order oscillatory response $u_1(t)$.
     
\end{document}